\documentclass[%
print,
 amsmath,amssymb,
 aps,twocolumn
]{revtex4}

\usepackage{times}
\usepackage{graphicx}
\usepackage{dcolumn}
\usepackage{bm}
\usepackage{dsfont}
\usepackage{mathrsfs}
\usepackage[landscape,papersize={297.1mm,210mm},left=1.9cm,right=1.6cm,top=2.7cm,bottom=2.8cm]{geometry}
\usepackage[                   
            pdfstartview=FitH,
            colorlinks, 
            pdfborder=001,   
            linkcolor=blue,
            anchorcolor=blue,
            citecolor=blue,
            urlcolor=blue
            ]{hyperref}
\usepackage{amsthm}
\makeatletter

\newcommand{\Rmnum}[1]{\expandafter\@slowromancap\romannumeral #1@}

\makeatother

\begin{document}
\title{Genuinely entangled subspaces and strongly nonlocal unextendible biseparable bases in four-partite systems}

\author{Huaqi Zhou$^{1,2}$}

\author{Ting Gao$^{3,4,5}$}
\email{gaoting@hebtu.edu.cn}

\author{Fengli Yan$^6$}
\email{flyan@hebtu.edu.cn}
\affiliation{$^1$ School of Mathematics and Science, Hebei GEO University, Shijiazhuang 052161, China\\
$^2$ Postdoctoral Research Station, School of Mathematical Sciences, Hebei Normal University, Shijiazhuang 050024, China\\
$^3$ School of Mathematical Sciences, Hebei Normal University, Shijiazhuang 050024, China \\
$^4$ Hebei Mathematics Research Center,  Hebei Normal University, Shijiazhuang 050024, China\\
$^5$ Hebei International Joint Research Center for Mathematics and Interdisciplinary Science, Hebei Normal University, Shijiazhuang 050024, China\\
$^6$ College of Physics, Hebei Key Laboratory of Photophysics Research and Application, Hebei Normal University, Shijiazhuang 050024, China}

\begin{abstract}
A set of orthogonal pure states is an unextendible biseparable basis (UBB), which means that its complementary subspace contains only genuinely entangled states. UBBs thus serve as an effective tool for constructing genuinely entangled subspaces. If every state within such a subspace exhibits distillable entanglement across all bipartitions, it becomes particularly advantageous for applications in quantum information. In this paper, we mainly conduct research on the 4-qudit quantum systems, where the local dimension $d$ is not less than 3. We present an approach for constructing UBB and prove that the UBB established in this way is strongly nonlocal. We build several genuinely entangled subspaces and demonstrate the distillability of the genuinely entangled subspaces across all bipartitions. In addition, we also describe the specific orthonormal basis for some genuinely entangled subspaces. These results will not only contribute to the development of quantum nonlocality theory, but also provide a crucial theoretical foundation for practical quantum information processing tasks.~\\


\end{abstract}

\maketitle

\section{Introduction}
Genuinely multipartite entanglement is one of the hallmark features of quantum mechanics \cite{Horodecki09,HGY12,GYV14}. It constitutes the strongest form of entanglement of a composite quantum system involving more than two spatially separated subsystem. A genuinely entangled state means that it possesses entanglement under every bipartition of the multipartite quantum system \cite{HGY12}. It has been identified as an object of both theoretical and practical significance and widely applied in fields such as quantum teleportation \cite{Bennett95,GYW05,GYL08,Luo19}, quantum key distribution \cite{Science283.2050,PRL67.661,PRL68.557}, and quantum state discrimination \cite{Rout19,ZGY22,ZGY23}. For this reason, genuinely entangled subspace (GES), which composed only of genuinely multipartite entangled states, gain recently considerable attention \cite{Baccari20,Demianowicz22,Demianowicz18,Agrawal19,Bhunia24}.

If we directly construct GES, it may increase the complexity of the problem. Recently, Demianowicz et al. \cite{Demianowicz18,Demianowicz22} explored the connection between unextendible product bases (UPBs) and GESs and provided a scheme to build the GESs by several classes of nonorthogonal UPBs for any number of parties and local dimensions. Here UPB expresses a set of orthogonal product states and the complementary subspace of its span contains no product state. Agrawal et al. \cite{Agrawal19} introduced the concept of unextendible biseparable bases (UBB) which is similar to UPB. The UBB is a set of pairwise orthogonal biseparable states and its complementary subspace contains no biseparable state. This implies that the complementary subspace of the proper subspace spanned by the UBB contains only genuinely entangled states. Naturally, there is an effective method to build the GES by researching the structure of UBB in some multipartite systems \cite{Bhunia24,Agrawal19}. Specifically, Agrawal et al. \cite{Agrawal19} provided explicit construction of UBBs for every 3-qudit ($d\geq 3$) quantum system and showed that the GES resulting from the UBB construction is indeed a bidistillable subspace. Bhunia et al. \cite{Bhunia24} derived a rule for constructing a class of UBB and the any state belonging to the corresponding GES is also distillable in every bipartition of system. Such entangled states can be distilled into maximally entangled states from certain (mixed) entangled states by only using local operations and classical communication (LOCC) \cite{Bennett96,Horodecki97,Horodecki98,Dur00}. They have extremely applicable value in practical information processing tasks.

It is worth noting that some UBBs have strong quantum nonlocality \cite{Bhunia24,Agrawal19} which is a property unique to the sets of orthogonal states and not limited by entangled systems \cite{Bennett99}. When classical information encoded in the sets of orthogonal states of a composite quantum system and all parties can only perform LOCC, strong quantum nonlocality of the set makes the information always be not decodable \cite{Halder19,ZGY25,HGY24,HGY25,Shi22}. This property can enhance the confidentiality of information and be widely applied to quantum data hiding \cite{Terhal01,Divincenzo02,Eggeling02,Matthews09} and quantum secret sharing \cite{Hillery99,Guo01,Hsu05,Markham08,JWang17}. In quantum systems $\mathbb{C}^{3}\otimes\mathbb{C}^{3}\otimes\mathbb{C}^{3}$ and $\mathbb{C}^{4}\otimes\mathbb{C}^{4}\otimes\mathbb{C}^{4}$, Halder et al. \cite{Halder19} have demonstrated strongly nonlocal orthogonal product sets. Yuan et al. \cite{Yuan20} presented the constructions of strongly nonlocal sets containing $6(d-1)^2$ orthogonal product states in $\mathbb{C}^{d}\otimes\mathbb{C}^{d}\otimes\mathbb{C}^{d}$ and $6d^2-8d+4$ orthogonal product states in $\mathbb{C}^{d}\otimes\mathbb{C}^{d}\otimes\mathbb{C}^{d+1}$, respectively. Here $d\geq 3$. We and He et al. \cite{ZGY23,He24} jointly provided a general construction of incomplete orthogonal product bases with strong nonlocality in any possible systems. Shi and Che et al. \cite{ShiL22,Che21} constructed strongly nonlocal unextendible product bases on the quantum system $\mathbb{C}^{d_{A}}\otimes\mathbb{C}^{d_{B}}\otimes\mathbb{C}^{d_{C}}$ ($d_{A},d_{B},d_{C}\geq 3$). There are many existing results concerning strongly nonlocal sets \cite{Walgate02,Johnston14,Zhang19,Shi20,Zhang20,ShiL21,LiM23,Bera25}. However, research on strongly nonlocal UBBs is still insufficient.

In this paper, we will study the strongly nonlocal unextendible biseparable bases and the genuinely entangled subspaces in four-partite systems by changing a family of orthogonal product set with strong quantum nonlocality. In Sec. \ref{Q2}, we introduce some required preliminary definitions and lemmas. In Sec. \ref{Q3}, we construct the strongly nonlocal UBB in 4-qutrit quantum system and provide a distillable GES in every bipartition. Next, in Sec. \ref{Q4}, we generalize the structure of the UBB to 4-qudit quantum systems and give the specific orthonormal basis of the GESs corresponding to the UBBs. Finally, we conclude  with a brief summary in Sec. \ref{Q5}.

\section{Preliminaries}\label{Q2}
In the quantum system $\mathcal{H}=\otimes_{k=1}^{n}\mathbb{C}^{d_{k}}$, a pure $n$-partite state $|\psi\rangle$ is called biseparable \cite{HGY12} if the $n$ parties can be partitioned into $2$ groups $A_{1}$ and $A_{2}$ such that the state can be written as a tensor product $|\psi\rangle_{A_{1}}\otimes|\psi\rangle_{A_{2}}$. Suppose $|\psi\rangle=\sum_{i,j}x_{ij}|i\rangle_{A_{1}}|j\rangle_{A_{2}}$ with a matrix $X=(x_{ij})$, where $\{|i\rangle_{A_{1}}\}$ and $\{|j\rangle_{A_{2}}\}$ are the computation bases of subsystems $\mathcal{H}_{A_{1}}$ and $\mathcal{H}_{A_{2}}$, respectively. Then, $\textrm{rank}(X)=1$ is a sufficient and necessary condition for $|\psi\rangle$ to be separable in the bipartition $A_{1}|A_{2}$ \cite{Bhunia24}. An $n$-partite mixed state $\rho$ is called biseparable if it can be written as the convex combination of biseparable pure states. The quantum states being not biseparable are genuinely entangled.

\emph{Definition 1} \cite{Agrawal19}. A set of pairwise orthogonal states $\{|\psi_{i}\rangle\}_{i=1}^m$ spanning a proper subspace of $\otimes_{j=1}^{n}\mathbb{C}^{d_{j}}$ is called an unextendible biseparable base (UBB), if all the states $|\psi_{i}\rangle$ are biseparable and its complementary subspace contains no biseparable state.

Genuinely entangled subspace (GES) represents a proper subspace with the property that all density matrices supported on it are genuinely entangled. It is obvious that the complementary subspace of the subspace spanned by a UBB is a GES.

\emph{Definition 2} \cite{Halder19}. A set of orthogonal quantum states on $\otimes_{j=1}^{n}\mathbb{C}^{d_{j}}$ with $n\geq 2$ and $d_{j}\geq 2$, is locally irreducible if it is not possible to distinguish one or more states from the set by orthogonality-preserving local measurements.

\emph{Definition 3} \cite{Halder19}. (Strongly quantum nonlocality) In multipartite system $\otimes_{j=1}^{n}\mathbb{C}^{d_{j}}$ with $n\geq 3$ and $d_{j}\geq 3$, a set of orthogonal quantum states is strongly nonlocal if it is locally irreducible in every bipartition.

A sufficient condition for a set of orthogonal states to be locally irreducible is that the orthogonality-preserving positive operator-valued measure (POVM) performed on every subsystem can only be trivial. Primarily, when the measurement is trivial, i.e., all the POVM elements of the measurement are proportional to the identity operator, it produces no information from the resulting postmeasurement states. Based on this, there is a sufficient condition of strongly quantum nonlocality as follows.

\emph{Lemma 1} \cite{ShiL22}. If $\overline{X_{i}}=\{1,2,\ldots,n\}\backslash \{i\}$ party can only perform a trivial orthogonality-preserving POVM for all $1\leq i\leq n$, then the set of orthogonal quantum states on $\mathcal{H}$ is of the strong nonlocality.

\emph{Lemma 2} \cite{Agrawal19}. For an $n$-dimensional subspace $\mathcal{S}_{\alpha\beta}$ of a bipartite Hilbert space $\mathbb{C}^{d_\alpha}\otimes\mathbb{C}^{d_\beta}$, if the projector $\mathbb{P}_{\alpha\beta}$ on $\mathcal{S}_{\alpha\beta}$ satisfies the condition $\mathcal{R}(\mathbb{P}_{\alpha\beta})<\max[\mathcal{R}(\mathbb{P}_{\alpha}),\mathcal{R}(\mathbb{P}_{\beta})]$, then all the rank-$n$ states supported on $\mathcal{S}_{\alpha\beta}$ are 1-distillable. Here $\mathcal{R}(\ast)$ expresses rank of a matrix and $\mathbb{P}_{\alpha(\beta)}:=\textrm{Tr}_{\beta(\alpha)}[\mathbb{P}_{\alpha\beta}]$.

For convenience, unless otherwise specified, the quantum states are not normalized in the following content. They differ from the normalized state by only a global coefficient.

\section{Strongly nonlocal UBB in 4-qutrit quantum system}\label{Q3}
Let $\mathcal{B}=\{|i_{1}i_{2}i_{3}i_{4}\rangle\}=\{|i_{1}\rangle|i_{2}\rangle|i_{3}\rangle|i_{4}\rangle\}$ ($i_{1},i_{2},i_{3},i_{4}\in \{0,1,2\}$) denote the computational basis of quantum system $\mathbb{C}^{3}\otimes\mathbb{C}^{3}\otimes\mathbb{C}^{3}\otimes\mathbb{C}^{3}$. In this system, the orthogonal product set $\mathcal{E}$ of Ref. \cite{ZGY23} is the union of the following subsets
\begin{equation}\label{ye4}
\begin{aligned}
&\mathcal{C}_{1}:=\{|\eta_{\pm}\rangle_{1}|0\rangle_{2}|0\rangle_{3}|0\rangle_{4}\},\\
&\mathcal{C}_{2}:=\{|0\rangle_{1}|\xi_{\pm}\rangle_{2}|2\rangle_{3}|2\rangle_{4}\},\\
&\mathcal{C}_{3}:=\{|0\rangle_{1}|0\rangle_{2}|\xi_{\pm}\rangle_{3}|2\rangle_{4}\},\\
&\mathcal{C}_{4}:=\{|0\rangle_{1}|0\rangle_{2}|0\rangle_{3}|\xi_{\pm}\rangle_{4}\},\\
&\mathcal{C}_{5}:=\{|\eta_{\pm}\rangle_{1}|\xi_{\pm}\rangle_{2}|\eta_{\pm}\rangle_{3}|0\rangle_{4}\},\\
&\mathcal{C}_{6}:=\{|\eta_{\pm}\rangle_{1}|\xi_{\pm}\rangle_{2}|2\rangle_{3}|\eta_{\pm}\rangle_{4}\},\\
&\mathcal{C}_{7}:=\{|\eta_{\pm}\rangle_{1}|0\rangle_{2}|\xi_{\pm}\rangle_{3}|\eta_{\pm}\rangle_{4}\},\\
&\mathcal{C}_{8}:=\{|0\rangle_{1}|\xi_{\pm}\rangle_{2}|\eta_{\pm}\rangle_{3}|\xi_{\pm}\rangle_{4}\},\\
&\mathcal{D}_{1}:=\{|\xi_{\pm}\rangle_{1}|2\rangle_{2}|2\rangle_{3}|2\rangle_{4}\},\\
&\mathcal{D}_{2}:=\{|2\rangle_{1}|\eta_{\pm}\rangle_{2}|0\rangle_{3}|0\rangle_{4}\},\\
&\mathcal{D}_{3}:=\{|2\rangle_{1}|2\rangle_{2}|\eta_{\pm}\rangle_{3}|0\rangle_{4}\},\\
&\mathcal{D}_{4}:=\{|2\rangle_{1}|2\rangle_{2}|2\rangle_{3}|\eta_{\pm}\rangle_{4}\},\\
&\mathcal{D}_{5}:=\{|\xi_{\pm}\rangle_{1}|\eta_{\pm}\rangle_{2}|\xi_{\pm}\rangle_{3}|2\rangle_{4}\},\\
&\mathcal{D}_{6}:=\{|\xi_{\pm}\rangle_{1}|\eta_{\pm}\rangle_{2}|0\rangle_{3}|\xi_{\pm}\rangle_{4}\},\\
&\mathcal{D}_{7}:=\{|\xi_{\pm}\rangle_{1}|2\rangle_{2}|\eta_{\pm}\rangle_{3}|\xi_{\pm}\rangle_{4}\},\\
&\mathcal{D}_{8}:=\{|2\rangle_{1}|\eta_{\pm}\rangle_{2}|\xi_{\pm}\rangle_{3}|\eta_{\pm}\rangle_{4}\}.\\
\end{aligned}
\end{equation}
Here $|\eta_{\pm}\rangle=|0\pm 1\rangle=|0\rangle\pm |1\rangle$ and $|\xi_{\pm}\rangle=|1\pm 2\rangle=|1\rangle\pm |2\rangle$. Subsets $\mathcal{C}_{1}$ and $\mathcal{D}_{2}$ are the bases spanned by the computational subbases $\{|0000\rangle,|1000\rangle\}$ and $\{|2000\rangle,|2100\rangle\}$, respectively. It is not difficult to find they have the same particle states $|0\rangle_{3}|0\rangle_{4}$ in the 3th and 4th subsystems. By removing the quantum states $|\eta_{+}\rangle_{1}|0\rangle_{2}|0\rangle_{3}|0\rangle_{4}$ of $\mathcal{C}_{1}$ and $|2\rangle_{1}|\eta_{+}\rangle_{2}|0\rangle_{3}|0\rangle_{4}$ of $\mathcal{D}_{2}$ and adding the quantum states $|0\rangle_{3}|0\rangle_{4}|00+10\pm(20+21)\rangle_{12}$, we get an orthogonal basis $\mathcal{U}_{1}$ of subspace generated by $\{|0000\rangle,|1000\rangle,|2000\rangle,|2100\rangle\}$ as follow
\begin{equation*}
\begin{aligned}
\mathcal{U}_{1}:=\{&|\psi_{1}\rangle^{C1}=|0-1\rangle_{1}|0\rangle_{2}|0\rangle_{3}|0\rangle_{4},\\
                     &|\psi_{1}\rangle^{D2}=|2\rangle_{1}|0-1\rangle_{2}|0\rangle_{3}|0\rangle_{4},\\
                     &|\psi_{\pm}\rangle^{U1}=|0\rangle_{3}|0\rangle_{4}|00+10\pm(20+21)\rangle_{12}\}.
\end{aligned}
\end{equation*}

Similarly, we provide the new subsets $\mathcal{U}_{2}$, $\mathcal{U}_{3}$, and $\mathcal{U}_{4}$ corresponding to $\mathcal{C}_{2}\cup\mathcal{D}_{1}$, $\mathcal{C}_{3}\cup\mathcal{C}_{4}$, and $\mathcal{D}_{3}\cup\mathcal{D}_{4}$ as follows
\begin{equation*}
\begin{aligned}
\mathcal{U}_{2}:=\{&|\psi_{1}\rangle^{C2}=|0\rangle_{1}|1-2\rangle_{2}|2\rangle_{3}|2\rangle_{4},\\
                   &|\psi_{1}\rangle^{D1}=|1-2\rangle_{1}|2\rangle_{2}|2\rangle_{3}|2\rangle_{4},\\
                   &|\psi_{\pm}\rangle^{U2}=|2\rangle_{3}|2\rangle_{4}|01+02\pm(12+22)\rangle_{12}\},\\
\mathcal{U}_{3}:=\{&|\psi_{1}\rangle^{C3}=|0\rangle_{1}|0\rangle_{2}|1-2\rangle_{3}|2\rangle_{4},\\
                   &|\psi_{1}\rangle^{C4}=|0\rangle_{1}|0\rangle_{2}|0\rangle_{3}|1-2\rangle_{4},\\
                   &|\psi_{\pm}\rangle^{U3}=|0\rangle_{1}|0\rangle_{2}|12+22\pm(01+02)\rangle_{34}\},\\
\mathcal{U}_{4}:=\{&|\psi_{1}\rangle^{D3}=|2\rangle_{1}|2\rangle_{2}|0-1\rangle_{3}|0\rangle_{4},\\
                   &|\psi_{1}\rangle^{D4}=|2\rangle_{1}|2\rangle_{2}|2\rangle_{3}|0-1\rangle_{4},\\
                   &|\psi_{\pm}\rangle^{U4}=|2\rangle_{1}|2\rangle_{2}|00+10\pm(20+21)\rangle_{34}\}.
\end{aligned}
\end{equation*}

For $\mathcal{C}_{5}$ and $\mathcal{C}_{8}$, they are the same particle states $|\xi_{\pm}\rangle_{2}|\eta_{\pm}\rangle_{3}$ in the 2th and 3th subsystems. We first remove the quantum states $|\eta_{+}\rangle_{1}|\xi_{+}\rangle_{2}|\eta_{+}\rangle_{3}|0\rangle_{4}$ and $|0\rangle_{1}|\xi_{+}\rangle_{2}|\eta_{+}\rangle_{3}|\xi_{+}\rangle_{4}$. Then, by adding the quantum states $|1+2\rangle_{2}|0+1\rangle_{3}|00+01\pm(10+20)\rangle_{41}$, we can construct a newly orthogonal basis $\mathcal{U}_{5}$ of subspace generated by $\mathcal{C}_{5}\cup\mathcal{C}_{8}$. Specifically,
\begin{equation*}
\begin{aligned}
\mathcal{U}_{5}:=\{&|\psi_{1}\rangle^{C5}=|\eta_{+}\rangle_{1}|\xi_{+}\rangle_{2}|\eta_{-}\rangle_{3}|0\rangle_{4},\\
                     &|\psi_{2}\rangle^{C5}=|\eta_{+}\rangle_{1}|\xi_{-}\rangle_{2}|\eta_{+}\rangle_{3}|0\rangle_{4},\\
                     &|\psi_{3}\rangle^{C5}=|\eta_{-}\rangle_{1}|\xi_{+}\rangle_{2}|\eta_{+}\rangle_{3}|0\rangle_{4},\\
                     &|\psi_{4}\rangle^{C5}=|\eta_{+}\rangle_{1}|\xi_{-}\rangle_{2}|\eta_{-}\rangle_{3}|0\rangle_{4},\\
                     &|\psi_{5}\rangle^{C5}=|\eta_{-}\rangle_{1}|\xi_{-}\rangle_{2}|\eta_{+}\rangle_{3}|0\rangle_{4},\\
                     &|\psi_{6}\rangle^{C5}=|\eta_{-}\rangle_{1}|\xi_{+}\rangle_{2}|\eta_{-}\rangle_{3}|0\rangle_{4},\\
                     &|\psi_{7}\rangle^{C5}=|\eta_{-}\rangle_{1}|\xi_{-}\rangle_{2}|\eta_{-}\rangle_{3}|0\rangle_{4},\\
                     &|\psi_{1}\rangle^{C8}=|0\rangle_{1}|\xi_{+}\rangle_{2}|\eta_{+}\rangle_{3}|\xi_{-}\rangle_{4},\\
                     &|\psi_{2}\rangle^{C8}=|0\rangle_{1}|\xi_{+}\rangle_{2}|\eta_{-}\rangle_{3}|\xi_{+}\rangle_{4},\\
                     &|\psi_{3}\rangle^{C8}=|0\rangle_{1}|\xi_{-}\rangle_{2}|\eta_{+}\rangle_{3}|\xi_{+}\rangle_{4},\\
                     &|\psi_{4}\rangle^{C8}=|0\rangle_{1}|\xi_{+}\rangle_{2}|\eta_{-}\rangle_{3}|\xi_{-}\rangle_{4},\\
                     &|\psi_{5}\rangle^{C8}=|0\rangle_{1}|\xi_{-}\rangle_{2}|\eta_{-}\rangle_{3}|\xi_{+}\rangle_{4},\\
                     &|\psi_{6}\rangle^{C8}=|0\rangle_{1}|\xi_{-}\rangle_{2}|\eta_{+}\rangle_{3}|\xi_{-}\rangle_{4},\\
                     &|\psi_{7}\rangle^{C8}=|0\rangle_{1}|\xi_{-}\rangle_{2}|\eta_{-}\rangle_{3}|\xi_{-}\rangle_{4},\\
                     &|\psi_{\pm}\rangle^{U5}=|\xi_{+}\rangle_{2}|\eta_{+}\rangle_{3}|00+01\pm(10+20)\rangle_{41}\}.
\end{aligned}
\end{equation*}

Similarly, corresponding to $\mathcal{C}_{6}\cup\mathcal{C}_{7}$, $\mathcal{D}_{5}\cup\mathcal{D}_{8}$, and $\mathcal{D}_{6}\cup\mathcal{D}_{7}$, we have
\begin{equation*}
\begin{aligned}
\mathcal{U}_{6}:=\{&|\psi_{1}\rangle^{C6}=|\eta_{+}\rangle_{1}|\xi_{+}\rangle_{2}|2\rangle_{3}|\eta_{-}\rangle_{4},\\
                     &|\psi_{2}\rangle^{C6}=|\eta_{+}\rangle_{1}|\xi_{-}\rangle_{2}|2\rangle_{3}|\eta_{+}\rangle_{4},\\
                     &|\psi_{3}\rangle^{C6}=|\eta_{-}\rangle_{1}|\xi_{+}\rangle_{2}|2\rangle_{3}|\eta_{+}\rangle_{4},\\
                     &|\psi_{4}\rangle^{C6}=|\eta_{+}\rangle_{1}|\xi_{-}\rangle_{2}|2\rangle_{3}|\eta_{-}\rangle_{4},\\
                     &|\psi_{5}\rangle^{C6}=|\eta_{-}\rangle_{1}|\xi_{-}\rangle_{2}|2\rangle_{3}|\eta_{+}\rangle_{4},\\
                     &|\psi_{6}\rangle^{C6}=|\eta_{-}\rangle_{1}|\xi_{+}\rangle_{2}|2\rangle_{3}|\eta_{-}\rangle_{4},\\
                     &|\psi_{7}\rangle^{C6}=|\eta_{-}\rangle_{1}|\xi_{-}\rangle_{2}|2\rangle_{3}|\eta_{-}\rangle_{4},\\
\end{aligned}
\end{equation*}

\begin{equation*}
\begin{aligned}
                     &|\psi_{1}\rangle^{C7}=|\eta_{+}\rangle_{1}|0\rangle_{2}|\xi_{+}\rangle_{3}|\eta_{-}\rangle_{4},\\
                     &|\psi_{2}\rangle^{C7}=|\eta_{+}\rangle_{1}|0\rangle_{2}|\xi_{-}\rangle_{3}|\eta_{+}\rangle_{4},\\
                     &|\psi_{3}\rangle^{C7}=|\eta_{-}\rangle_{1}|0\rangle_{2}|\xi_{+}\rangle_{3}|\eta_{+}\rangle_{4},\\
                     &|\psi_{4}\rangle^{C7}=|\eta_{+}\rangle_{1}|0\rangle_{2}|\xi_{-}\rangle_{3}|\eta_{-}\rangle_{4},\\
                     &|\psi_{5}\rangle^{C7}=|\eta_{-}\rangle_{1}|0\rangle_{2}|\xi_{-}\rangle_{3}|\eta_{+}\rangle_{4},\\
                     &|\psi_{6}\rangle^{C7}=|\eta_{-}\rangle_{1}|0\rangle_{2}|\xi_{+}\rangle_{3}|\eta_{-}\rangle_{4},\\
                     &|\psi_{7}\rangle^{C7}=|\eta_{-}\rangle_{1}|0\rangle_{2}|\xi_{-}\rangle_{3}|\eta_{-}\rangle_{4},\\
                     &|\psi_{\pm}\rangle^{U6}=|\eta_{+}\rangle_{4}|\eta_{+}\rangle_{1}|12+22\pm(01+02)\rangle_{23}\},\\
\mathcal{U}_{7}:=\{&|\psi_{1}\rangle^{D5}=|\xi_{+}\rangle_{1}|\eta_{+}\rangle_{2}|\xi_{-}\rangle_{3}|2\rangle_{4},\\
                     &|\psi_{2}\rangle^{D5}=|\xi_{+}\rangle_{1}|\eta_{-}\rangle_{2}|\xi_{+}\rangle_{3}|2\rangle_{4},\\
                     &|\psi_{3}\rangle^{D5}=|\xi_{-}\rangle_{1}|\eta_{+}\rangle_{2}|\xi_{+}\rangle_{3}|2\rangle_{4},\\
                     &|\psi_{4}\rangle^{D5}=|\xi_{+}\rangle_{1}|\eta_{-}\rangle_{2}|\xi_{-}\rangle_{3}|2\rangle_{4},\\
                     &|\psi_{5}\rangle^{D5}=|\xi_{-}\rangle_{1}|\eta_{-}\rangle_{2}|\xi_{+}\rangle_{3}|2\rangle_{4},\\
                     &|\psi_{6}\rangle^{D5}=|\xi_{-}\rangle_{1}|\eta_{+}\rangle_{2}|\xi_{-}\rangle_{3}|2\rangle_{4},\\
                     &|\psi_{7}\rangle^{D5}=|\xi_{-}\rangle_{1}|\eta_{-}\rangle_{2}|\xi_{-}\rangle_{3}|2\rangle_{4},\\
                     &|\psi_{1}\rangle^{D8}=|2\rangle_{1}|\eta_{+}\rangle_{2}|\xi_{+}\rangle_{3}|\eta_{-}\rangle_{4},\\
                     &|\psi_{2}\rangle^{D8}=|2\rangle_{1}|\eta_{+}\rangle_{2}|\xi_{-}\rangle_{3}|\eta_{+}\rangle_{4},\\
                     &|\psi_{3}\rangle^{D8}=|2\rangle_{1}|\eta_{-}\rangle_{2}|\xi_{+}\rangle_{3}|\eta_{+}\rangle_{4},\\
                     &|\psi_{4}\rangle^{D8}=|2\rangle_{1}|\eta_{+}\rangle_{2}|\xi_{-}\rangle_{3}|\eta_{-}\rangle_{4},\\
                     &|\psi_{5}\rangle^{D8}=|2\rangle_{1}|\eta_{-}\rangle_{2}|\xi_{-}\rangle_{3}|\eta_{+}\rangle_{4},\\
                     &|\psi_{6}\rangle^{D8}=|2\rangle_{1}|\eta_{-}\rangle_{2}|\xi_{+}\rangle_{3}|\eta_{-}\rangle_{4},\\
                     &|\psi_{7}\rangle^{D8}=|2\rangle_{1}|\eta_{-}\rangle_{2}|\xi_{-}\rangle_{3}|\eta_{-}\rangle_{4},\\
                     &|\psi_{\pm}\rangle^{U7}=|\eta_{+}\rangle_{2}|\xi_{+}\rangle_{3}|21+22\pm(02+12)\rangle_{41}\},\\
\mathcal{U}_{8}:=\{&|\psi_{1}\rangle^{D6}=|\xi_{+}\rangle_{1}|\eta_{+}\rangle_{2}|0\rangle_{3}|\xi_{-}\rangle_{4},\\
                     &|\psi_{2}\rangle^{D6}=|\xi_{+}\rangle_{1}|\eta_{-}\rangle_{2}|0\rangle_{3}|\xi_{+}\rangle_{4},\\
                     &|\psi_{3}\rangle^{D6}=|\xi_{-}\rangle_{1}|\eta_{+}\rangle_{2}|0\rangle_{3}|\xi_{+}\rangle_{4},\\
                     &|\psi_{4}\rangle^{D6}=|\xi_{+}\rangle_{1}|\eta_{-}\rangle_{2}|0\rangle_{3}|\xi_{-}\rangle_{4},\\
                     &|\psi_{5}\rangle^{D6}=|\xi_{-}\rangle_{1}|\eta_{-}\rangle_{2}|0\rangle_{3}|\xi_{+}\rangle_{4},\\
                     &|\psi_{6}\rangle^{D6}=|\xi_{-}\rangle_{1}|\eta_{+}\rangle_{2}|0\rangle_{3}|\xi_{-}\rangle_{4},\\
                     &|\psi_{7}\rangle^{D6}=|\xi_{-}\rangle_{1}|\eta_{-}\rangle_{2}|0\rangle_{3}|\xi_{-}\rangle_{4},\\
                     &|\psi_{1}\rangle^{D7}=|\xi_{+}\rangle_{1}|2\rangle_{2}|\eta_{+}\rangle_{3}|\xi_{-}\rangle_{4},\\
                     &|\psi_{2}\rangle^{D7}=|\xi_{+}\rangle_{1}|2\rangle_{2}|\eta_{-}\rangle_{3}|\xi_{+}\rangle_{4},\\
                     &|\psi_{3}\rangle^{D7}=|\xi_{-}\rangle_{1}|2\rangle_{2}|\eta_{+}\rangle_{3}|\xi_{+}\rangle_{4},\\
                     &|\psi_{4}\rangle^{D7}=|\xi_{+}\rangle_{1}|2\rangle_{2}|\eta_{-}\rangle_{3}|\xi_{-}\rangle_{4},\\
                     &|\psi_{5}\rangle^{D7}=|\xi_{-}\rangle_{1}|2\rangle_{2}|\eta_{-}\rangle_{3}|\xi_{+}\rangle_{4},\\
                     &|\psi_{6}\rangle^{D7}=|\xi_{-}\rangle_{1}|2\rangle_{2}|\eta_{+}\rangle_{3}|\xi_{-}\rangle_{4},\\
                     &|\psi_{7}\rangle^{D7}=|\xi_{-}\rangle_{1}|2\rangle_{2}|\eta_{-}\rangle_{3}|\xi_{-}\rangle_{4},\\
                     &|\psi_{\pm}\rangle^{U8}=|\xi_{+}\rangle_{4}|\xi_{+}\rangle_{1}|00+10\pm(20+21)\rangle_{23}\}.\\
\end{aligned}
\end{equation*}

Let $\mathcal{U}=\big(\bigcup_{i=1}^{8}\mathcal{U}_{i}^{-}\big)\bigcup\{|S\rangle\}$, where $\mathcal{U}_{i}^{-}= \mathcal{U}_{i}\setminus \{|\psi_{+}\rangle^{Ui}\}$ and $|S\rangle=|0+1+2\rangle_{1}|0+1+2\rangle_{2}|0+1+2\rangle_{3}|0+1+2\rangle_{4}$. $|S\rangle$ is a stopper state and orthogonal to all quantum states in $\bigcup_{i=1}^{8}\mathcal{U}_{i}^{-}$. Moreover, the sets $\mathcal{U}_{i}$ ($i=1,2,\ldots,8$) are mutually orthogonal because they are the bases of pairwise orthogonal subspaces. So, $\mathcal{U}$ is a set of orthogonal states. About this set, we have the following conclusion.

\emph{Theorem 1}. In $\mathbb{C}^{3}\otimes\mathbb{C}^{3}\otimes\mathbb{C}^{3}\otimes\mathbb{C}^{3}$ quantum system, the set $\mathcal{U}$ is an unextendible biseparable base (UBB). Moreover, an orthonormal basis of its complementary subspace can be expressed as
\begin{equation}
\begin{aligned}
&|G_1\rangle=\frac{1}{\sqrt{2}}\left(|\widetilde{\psi_{+}}\rangle^{U1}-|\widetilde{\psi_{+}}\rangle^{U2}\right),\\
&|G_2\rangle=\frac{1}{\sqrt{2}}\left(|\widetilde{\psi_{+}}\rangle^{U3}-|\widetilde{\psi_{+}}\rangle^{U4}\right),\\
&|G_3\rangle=\frac{1}{2}\left(|\widetilde{\psi_{+}}\rangle^{U1}+|\widetilde{\psi_{+}}\rangle^{U2}-|\widetilde{\psi_{+}}\rangle^{U3}-|\widetilde{\psi_{+}}\rangle^{U4}\right),\\
&|G_4\rangle=\frac{1}{\sqrt{2}}\left(|\widetilde{\psi_{+}}\rangle^{U5}-|\widetilde{\psi_{+}}\rangle^{U6}\right),\\
&|G_5\rangle=\frac{1}{\sqrt{2}}\left(|\widetilde{\psi_{+}}\rangle^{U7}-|\widetilde{\psi_{+}}\rangle^{U8}\right),\\
&|G_6\rangle=\frac{1}{2}\left(|\widetilde{\psi_{+}}\rangle^{U5}+|\widetilde{\psi_{+}}\rangle^{U6}-|\widetilde{\psi_{+}}\rangle^{U7}-|\widetilde{\psi_{+}}\rangle^{U8}\right),\\
&|G_7\rangle=\frac{1}{\sqrt{5}}\left(\sum_{i=1}^4|\widetilde{\psi_{+}}\rangle^{Ui}\right)-\frac{1}{2\sqrt{5}}\left(\sum_{i=5}^8|\widetilde{\psi_{+}}\rangle^{Ui}\right),\\
\end{aligned}
\end{equation}\\
\begin{equation*}
\begin{aligned}
&|G_8\rangle=\frac{1}{18\sqrt{5}}\left(\sum_{i=1}^4|\widetilde{\psi_{+}}\rangle^{Ui}\right)+\frac{1}{9\sqrt{5}}\left(\sum_{i=5}^8|\widetilde{\psi_{+}}\rangle^{Ui}\right)\\ &~~~~~~~~~~~~-\frac{4\sqrt{5}}{9}|1111\rangle.\\
\end{aligned}
\end{equation*}
Here $|\widetilde{\psi_{+}}\rangle^{Ui}$ is the normalized state of $|\psi_{+}\rangle^{Ui}$ with $i=1,2,\ldots,8$.

$Proof$. The set $\mathcal{U}$ is obtained by changing the orthogonal product set $\mathcal{E}$ (\ref{ye4}). It is obvious that the states belonging to set $\mathcal{U}$ are biseparable and pairwise orthogonal. Let $\mathcal{H}_{\mathcal{U}}$ denote the proper subspace spanned by set $\mathcal{U}$ and $\mathcal{H}_{\mathcal{U}}^{\bot}$ be its complementary subspace. To prove that $\mathcal{U}$ is a UBB, we only need to explain that every state in $\mathcal{H}_{\mathcal{U}}^{\bot}$ is a genuinely entangled state, i.e. the state is inseparable in all bipartitions. Because the states $|1111\rangle$ and $|\psi_{+}\rangle^{Ui}$ $(i=1,2,\ldots,8)$ are not orthogonal to $|S\rangle$ but are orthogonal to all states in $\mathcal{U}\setminus|S\rangle$, the any state $|\phi\rangle\in\mathcal{H}_{\mathcal{U}}^{\bot}$ can be written as a linear combination $a|\psi_{+}\rangle^{U1}+b|\psi_{+}\rangle^{U2}+c|\psi_{+}\rangle^{U3}+d|\psi_{+}\rangle^{U4}+e|\psi_{+}\rangle^{U5}+ f|\psi_{+}\rangle^{U6}+g|\psi_{+}\rangle^{U7}+h|\psi_{+}\rangle^{U8}+k|1111\rangle$. Here the coefficients have at least two non-zero values and such that $\langle \phi|S\rangle=0$.
~\\
\begin{widetext}
Assume $|\phi\rangle$ is separable in $1|234$ bipartition. Then, it corresponds to a matrix
\begin{equation*}
\begin{aligned}
M_1=\left(\begin{array}{ccccccccccccccccccccccccccc}
a & c & c & f & f & c & f & f & c & e & e & e & e & e & e & f & f & b & e & e & e & e & e & e & f & f & b \\
a & h & h & f & f & g & f & f & g & e & h & h & e & k & g & f & f & g & e & h & h & e & h & h & f & f & b \\
a & h & h & g & g & g & g & g & g & a & h & h & g & g & g & g & g & g & d & h & h & d & h & h & d & d & b \\
\end{array}
\right),
\end{aligned}
\end{equation*}
and $\textrm{rank}(M_1)=1$. Combine with the condition of at least two non-zero coefficients, it can be deduce that $a=b=c=d=e=f=g=h=k\neq 0$. Meanwhile, in $1|234$ bipartition, $|S\rangle$ corresponds to the matrix
\begin{equation*}
\begin{aligned}
J=\left(\begin{array}{ccccccccccccccccccccccccccc}
1 & 1 & 1 & 1 & 1 & 1 & 1 & 1 & 1 & 1 & 1 & 1 & 1 & 1 & 1 & 1 & 1 & 1 & 1 & 1 & 1 & 1 & 1 & 1 & 1 & 1 & 1 \\
1 & 1 & 1 & 1 & 1 & 1 & 1 & 1 & 1 & 1 & 1 & 1 & 1 & 1 & 1 & 1 & 1 & 1 & 1 & 1 & 1 & 1 & 1 & 1 & 1 & 1 & 1 \\
1 & 1 & 1 & 1 & 1 & 1 & 1 & 1 & 1 & 1 & 1 & 1 & 1 & 1 & 1 & 1 & 1 & 1 & 1 & 1 & 1 & 1 & 1 & 1 & 1 & 1 & 1 \\
\end{array}
\right).
\end{aligned}
\end{equation*}
The inner product $\langle \phi|S\rangle=\textrm{Tr}(M_1^{\dagger}J)\neq 0$. This is contradictory to the orthogonality between $|\phi\rangle$ and $|S\rangle$. So, $|\phi\rangle$ is not separable in $1|234$ bipartition.
\end{widetext}

In other bipartitions $2|341$, $3|412$, $4|123$, $12|34$, $13|24$, and $14|23$, the matrix representations of $|\phi\rangle$ are provided in Appendix \ref{A}. Similarly, we can get same conclusion.  This means that the complementary subspace $\mathcal{H}_{\mathcal{U}}^{\bot}$ contains only genuinely entangled states. Therefore, the set $\mathcal{U}$ is an unextendible biseparable base (UBB).

On the other point, it is easy to know that the set $\{|G_i\rangle\}$ is an orthonormal basis of complementary subspace $\mathcal{H}_{\mathcal{U}}^{\bot}$, since $\textrm{dim}(\mathcal{H}_{\mathcal{U}}^{\bot})=8$ and the eight orthonormal states $|G_i\rangle$ $(i=1,2,\ldots,8)$ are all the linear combinations that meet the above two conditions.
~ \hfill $\square$

\emph{Theorem 2}. The set $\mathcal{U}$ on $\mathbb{C}^{3}\otimes\mathbb{C}^{3}\otimes\mathbb{C}^{3}\otimes\mathbb{C}^{3}$ quantum system has the strongest quantum nonlocality.

$Proof$. We first consider the case in bipartition $X_{1}|X_{234}$. Here $X_{1}=\{1\}$ and $X_{234}=\{2,3,4\}$ express the union of subsystems of $\mathbb{C}_{1}^{3}\otimes\mathbb{C}_{2}^{3}\otimes\mathbb{C}_{3}^{3}\otimes\mathbb{C}_{4}^{3}$. Fig. \ref{Tu2} shows the structure of set $\mathcal{U}$ under this bipartition. We only need to prove that all orthogonality-preserving POVMs performed on the party $X_{234}$ are trivial. Suppose $M=(m_{kl})$ be an any orthogonality-preserving POVM element in the computation basis $\mathcal{B}^{\{234\}}=\{|i_{2}i_{3}i_{4}\rangle\}_{i_{2},i_{3},i_{4}=0}^2$, where $k,l\in\{i_{2}i_{3}i_{4}~|~i_{2},i_{3},i_{4}=0,1,2\}$. Then $\langle \phi_{1}|\mathbb{I}\otimes M|\phi_{2}\rangle=0$ for any two quantum states $|\phi_{1}\rangle$, $|\phi_{2}\rangle$ belonging to set $\mathcal{U}$.

Let $\mathcal{P}_{j}^{(234)}=\{|i_{2}i_{3}i_{4}\rangle~| ~\langle i_{1}i_{2}i_{3}i_{4}|\psi\rangle\neq 0,~|i_{1}i_{2}i_{3}i_{4}\rangle\in\mathcal{B},~|\psi\rangle\in\mathcal{P}_{j}\}$ for $\mathcal{P}\in \{\mathcal{C},\mathcal{D}\}$ and $j=1,2,\ldots,8$. Define
\begin{equation}
_{\mathcal{P}_{j_{1}}}M_{\mathcal{P}_{j_{2}}}=\sum_{|k\rangle\in\mathcal{P}_{j_{1}}^{(234)}}\sum_{|l\rangle\in\mathcal{P}_{j_{2}}^{(234)}} m_{kl}|k\rangle\langle l|
\end{equation}
\begin{widetext}
~
\begin{figure}[h]
\centering
\includegraphics[width=0.95\textwidth]{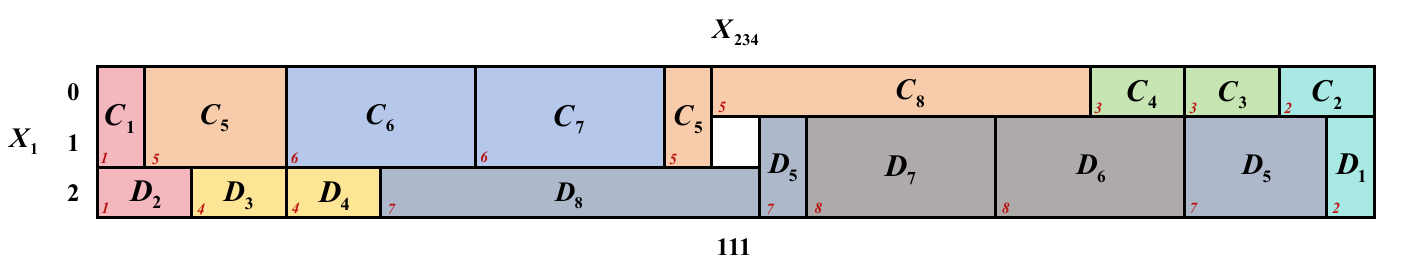}
\caption{This is the structure of set $\mathcal{U}$ under bipartition $X_{1}|X_{234}$. The regions denoted as $\mathcal{C}_{j}$ and $\mathcal{D}_{j}$ with $j=1,\ldots,8$ correspond to the subsets of Eq. (\ref{ye4}), respectively. Each colored area is marked with a red number in the left lower corner. The union of the regions with the same color and red number $j$ corresponds the subsets $\mathcal{U}_{j}^{-}$ of set $\mathcal{U}$. \label{Tu2}}
\end{figure}
\end{widetext}
as the subblock of matrix $M$ corresponding to subsets $\mathcal{P}_{j_{1}}$ and $\mathcal{P}_{j_{2}}$. For convenience, we use $M_{\mathcal{P}_{j_{1}}}$ to express $_{\mathcal{P}_{j_{1}}}M_{\mathcal{P}_{j_{1}}}$. Here $_{\mathcal{P}_{j_{1}}}M_{\mathcal{P}_{j_{2}}}$ is equal to the Hermitian conjugate of $_{\mathcal{P}_{j_{2}}}M_{\mathcal{P}_{j_{1}}}$ due to the property of POVM element. Choose $|\psi_{1}\rangle^{C1},|\psi_{1}\rangle^{D1}\in \mathcal{U}$, we can deduce $m_{000,222}=0$, i.e. $_{\mathcal{C}_{1}}M_{\mathcal{D}_{1}}={_{\mathcal{D}_{1}}M_{\mathcal{C}_{1}}}=\textbf{0}$. At this moment, consider $|\psi_{1}\rangle^{C1}$ and  $|\psi_{1}\rangle^{C2}$, we have $_{\mathcal{C}_{1}}M_{\mathcal{C}_{2}}=\textbf{0}$. According to the Block Zeros Lemma of Ref. \cite{ShiL22}, it is easy to know that the calculate between $|\psi_{1}\rangle^{C1}$ and $\{|\psi_{3}\rangle^{C5},|\psi_{5}\rangle^{C5},|\psi_{6}\rangle^{C5},|\psi_{7}\rangle^{C5}\}$ has the same effect as the calculate between $\mathcal{C}_{1}$ and $\mathcal{C}_{5}$. Hence, $_{\mathcal{C}_{1}}M_{\mathcal{C}_{5}}=\textbf{0}$. Then, consider $|\psi_{1}\rangle^{C1}$ and $\mathcal{U}_{5}^{-}$, we get $_{\mathcal{C}_{1}}M_{\mathcal{C}_{8}}=\textbf{0}$. Combining with the construction (Fig. \ref{Tu2}) of set $\mathcal{U}$, similar results are shown below.

\begin{equation*}
\begin{aligned}
&\mathcal{U}_{1}^{-},\mathcal{U}_{6}^{-} && \Rightarrow && _{\mathcal{C}_{1}}M_{\mathcal{C}_{6}}={_{\mathcal{C}_{1}}M_{\mathcal{C}_{7}}}=\textbf{0},\\
&\mathcal{U}_{1}^{-},\mathcal{U}_{7}^{-} && \Rightarrow && _{\mathcal{C}_{1}}M_{\mathcal{D}_{5}}=\textbf{0},\\
&\mathcal{U}_{1}^{-},\mathcal{U}_{8}^{-} && \Rightarrow && _{\mathcal{C}_{1}}M_{\mathcal{D}_{6}}={_{\mathcal{C}_{1}}M_{\mathcal{D}_{7}}}=\textbf{0},\\
&\mathcal{U}_{5}^{-},\mathcal{U}_{6}^{-} && \Rightarrow && _{\mathcal{C}_{5}}M_{\mathcal{C}_{6}}={_{\mathcal{C}_{5}}M_{\mathcal{C}_{7}}}=\\
&&&&&_{\mathcal{C}_{8}}M_{\mathcal{C}_{6}}={_{\mathcal{C}_{8}}M_{\mathcal{C}_{7}}}=\textbf{0},\\
&\mathcal{U}_{5}^{-},\mathcal{U}_{7}^{-} && \Rightarrow && _{\mathcal{C}_{5}}M_{\mathcal{D}_{5}}=\textbf{0},\\
&\mathcal{U}_{5}^{-},\mathcal{U}_{8}^{-} && \Rightarrow && _{\mathcal{C}_{5}}M_{\mathcal{D}_{6}}={_{\mathcal{C}_{5}}M_{\mathcal{D}_{7}}}=\textbf{0},\\
&\mathcal{U}_{5}^{-},\mathcal{U}_{2}^{-} && \Rightarrow && _{\mathcal{C}_{5}}M_{\mathcal{D}_{1}}=\textbf{0},\\
&\mathcal{U}_{6}^{-},\mathcal{U}_{6}^{-} && \Rightarrow && _{\mathcal{C}_{6}}M_{\mathcal{C}_{7}}=\textbf{0},\\
&\mathcal{U}_{6}^{-},\mathcal{U}_{7}^{-} && \Rightarrow && _{\mathcal{C}_{6}}M_{\mathcal{D}_{5}}={_{\mathcal{C}_{7}}M_{\mathcal{D}_{5}}}=\textbf{0},\\
&\mathcal{U}_{6}^{-},\mathcal{U}_{8}^{-} && \Rightarrow && _{\mathcal{C}_{6}}M_{\mathcal{D}_{6}}={_{\mathcal{C}_{6}}M_{\mathcal{D}_{7}}}=\\
&&&&&_{\mathcal{C}_{7}}M_{\mathcal{D}_{6}}= {_{\mathcal{C}_{7}}M_{\mathcal{D}_{7}}}=\textbf{0},\\
&\mathcal{U}_{6}^{-},\mathcal{U}_{2}^{-} && \Rightarrow && _{\mathcal{C}_{6}}M_{\mathcal{D}_{1}}={_{\mathcal{C}_{7}}M_{\mathcal{D}_{1}}}=\textbf{0},\\
&\mathcal{U}_{7}^{-},\mathcal{U}_{8}^{-} && \Rightarrow && _{\mathcal{D}_{5}}M_{\mathcal{D}_{6}}={_{\mathcal{D}_{5}}M_{\mathcal{D}_{7}}}=\\
&&&&&_{\mathcal{D}_{8}}M_{\mathcal{D}_{6}}= {_{\mathcal{D}_{8}}M_{\mathcal{D}_{7}}}=\textbf{0},\\
&\mathcal{U}_{7}^{-},\mathcal{U}_{2}^{-} && \Rightarrow && _{\mathcal{D}_{5}}M_{\mathcal{D}_{1}}={_{\mathcal{D}_{8}}M_{\mathcal{D}_{1}}}=\textbf{0},\\
&\mathcal{U}_{8}^{-},\mathcal{U}_{8}^{-} && \Rightarrow && _{\mathcal{D}_{6}}M_{\mathcal{D}_{7}}=\textbf{0},\\
&\mathcal{U}_{8}^{-},\mathcal{U}_{2}^{-} && \Rightarrow && _{\mathcal{D}_{6}}M_{\mathcal{D}_{1}}={_{\mathcal{D}_{7}}M_{\mathcal{D}_{1}}}=\textbf{0}.
\end{aligned}
\end{equation*}

\begin{figure}[h]
\centering
\includegraphics[width=0.4\textwidth]{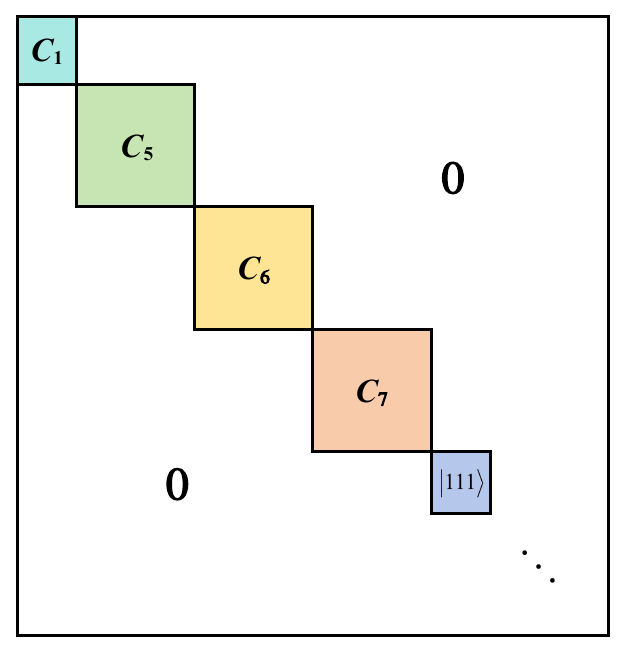}
\caption{Divide the matrix $M$ into blocks based on $|111\rangle$ and $\mathcal{P}_{j}^{(234)}$ $(\mathcal{P}=\mathcal{C},\mathcal{D}$ and $j=1,5,6,7)$. The off-diagonal subblocks are all zero matrix. In figure, $\mathcal{P}_{j}$ represents the subblock $M_{\mathcal{P}_{j}}$. \label{Tu1}}
\end{figure}

Moreover, in $\mathcal{U}_{5}^{-}$ and $\mathcal{U}_{7}^{-}$, $_{\mathcal{C}_{5}}M_{\mathcal{C}_{8}}=\textbf{0}$ and $_{\mathcal{D}_{5}}M_{\mathcal{D}_{8}}=\textbf{0}$ because $_{\mathcal{C}_{5}}M_{\mathcal{D}_{7}}=\textbf{0}$ and $_{\mathcal{D}_{5}}M_{\mathcal{C}_{7}}=\textbf{0}$, respectively. Thus, all elements of the operator $M$ are zero except for $m_{000,000}$, $m_{111,111}$, $m_{222,222}$, and the elements in $M_{\mathcal{P}_{j}}$ $(\mathcal{P}=\mathcal{C},\mathcal{D}$ and $j=5,6,7)$, as shown in the Fig. \ref{Tu1}.

Next, we will discuss whether $M$ is proportional to identity operator $\mathbb{I}$. From $_{\mathcal{C}_{1}}M_{\mathcal{C}_{6}}={_{\mathcal{C}_{5}}M_{\mathcal{C}_{6}}}=\textbf{0}$, we have $_{\mathcal{D}_{2}}M_{\mathcal{D}_{4}}=\textbf{0}$. So, in $\mathcal{U}_{1}^{-}$ and $\mathcal{U}_{4}^{-}$, we get $_{\mathcal{D}_{2}}M_{\mathcal{D}_{3}}=\textbf{0}$. Furthermore, we can infer that $_{\mathcal{D}_{2}}M_{\mathcal{D}_{8}}=\textbf{0}$ from $_{\mathcal{C}_{1}}M_{\mathcal{C}_{5}}={_{\mathcal{C}_{1}}M_{\mathcal{C}_{6}}}={_{\mathcal{C}_{1}}M_{\mathcal{C}_{7}}}={_{\mathcal{C}_{1}}M_{\mathcal{C}_{8}}} ={_{\mathcal{C}_{5}}M_{\mathcal{C}_{6}}}={_{\mathcal{C}_{5}}M_{\mathcal{C}_{7}}}={_{\mathcal{C}_{5}}M_{\mathcal{C}_{8}}}=\textbf{0}$. By using the Block Trivial Lemma of Ref. \cite{ShiL22}, we deduce $M_{\mathcal{C}_{5}}\propto \mathbb{I}$. Analogously, $M_{\mathcal{D}_{2}}\propto \mathbb{I}$ and $M_{\mathcal{D}_{8}}\propto \mathbb{I}$ because $_{\mathcal{C}_{1}}M_{\mathcal{C}_{5}}=\textbf{0}$ and $_{\mathcal{C}_{5}}M_{\mathcal{C}_{6}}={_{\mathcal{C}_{5}}M_{\mathcal{C}_{7}}}={_{\mathcal{C}_{5}}M_{\mathcal{C}_{8}}}=\textbf{0}$, respectively. According to $M_{\mathcal{C}_{5}}\propto \mathbb{I}$ and $_{\mathcal{C}_{6}}M_{\mathcal{C}_{5}}={_{\mathcal{C}_{7}}M_{\mathcal{C}_{5}}}={_{\mathcal{C}_{8}}M_{\mathcal{C}_{5}}}=\textbf{0}$, we have $_{\mathcal{D}_{8}}M_{\mathcal{D}_{3}}=\textbf{0}$. Moreover, $_{\mathcal{D}_{5}}M_{\mathcal{D}_{3}}={_{\mathcal{D}_{5}}M_{\mathcal{D}_{4}}}=\textbf{0}$ due to $_{\mathcal{D}_{5}}M_{\mathcal{C}_{5}}={_{\mathcal{D}_{5}}M_{\mathcal{C}_{6}}}=\textbf{0}$. So, in $\mathcal{U}_{4}^{-}$ and $\mathcal{U}_{7}^{-}$, $_{\mathcal{D}_{8}}M_{\mathcal{D}_{4}}=\textbf{0}$. Also, because $M_{\mathcal{D}_{8}}\propto \mathbb{I}$, we get $M_{\mathcal{C}_{6}}\propto \mathbb{I}$. Due to the symmetry of the structure of set $\mathcal{U}$, we can similarly derive that $M_{\mathcal{D}_{5}}\propto \mathbb{I}$, $M_{\mathcal{C}_{2}}\propto \mathbb{I}$, $M_{\mathcal{C}_{8}}\propto \mathbb{I}$, $M_{\mathcal{D}_{6}}\propto \mathbb{I}$. It is not difficult to find that the off-diagonal elements of $M$ are all zero owing to $\mathcal{C}_{7}^{(234)}\subset \mathcal{D}_{8}^{(234)}$ and $\mathcal{D}_{7}^{(234)}\subset \mathcal{C}_{8}^{(234)}$. Moreover, because $\mathcal{D}_{2}^{(234)}\cap \mathcal{C}_{5}^{(234)}\neq \emptyset$, $\mathcal{C}_{5}^{(234)}\cap \mathcal{D}_{8}^{(234)}\neq \emptyset$, $\mathcal{D}_{8}^{(234)}\cap \mathcal{C}_{6}^{(234)}\neq \emptyset$, $\mathcal{D}_{8}^{(234)}\cap \mathcal{C}_{8}^{(234)}\neq \emptyset$, $\mathcal{C}_{8}^{(234)}\cap \mathcal{D}_{6}^{(234)}\neq \emptyset$, $\mathcal{C}_{8}^{(234)}\cap \mathcal{D}_{5}^{(234)}\neq \emptyset$, $\mathcal{D}_{5}^{(234)}\cap \mathcal{C}_{2}^{(234)}\neq \emptyset$, and $\bigcup_{\mathcal{P},j}\mathcal{P}_{j}^{(234)}=\mathcal{B}^{\{234\}}$ $(\mathcal{P}=\mathcal{C},\mathcal{D}$ and $j=2,5,6,8)$, we can know that the diagonal elements of operator $M$ are all equal. That is, $M\propto \mathbb{I}$. For the other three bipartitions $X_{2}|X_{341}$, $X_{3}|X_{412}$, and $X_{4}|X_{123}$, we can get similar results based on the structural characteristics of set $\mathcal{U}$. Therefore, the set $\mathcal{U}$ has the strongest quantum nonlocality.  ~ \hfill $\square$

Let $\mathcal{H}_{\mathcal{S}}$ represent the subspace spanned by corresponding set $\mathcal{S}$. The orthogonal complementary subspace $\mathcal{H}_{\mathcal{U}}^{\bot}$ of $\mathcal{H}_{\mathcal{U}}$ is also equal to $\mathcal{H}_{\mathcal{G}}\backslash \mathcal{H}_{\{|S\rangle\}}$, i.e., the complementary subspace of $\mathcal{H}_{\{|S\rangle\}}$ in $\mathcal{H}_{\mathcal{G}}$ for $\mathcal{G}=\{|\psi_{+}\rangle^{Ui}\}_{i=1}^{8}\cup\{|1111\rangle\}$. Define $\mathcal{G}_{1}=\{|\psi_{+}\rangle^{Ui}\}_{i=1}^{7}$. Obviously, $\mathcal{H}_{\mathcal{G}_{1}}\backslash \mathcal{H}_{\{|S\rangle\}}$ is a proper subspace of $\mathcal{H}_{\mathcal{U}}^{\bot}$. We have proved that $\mathcal{H}_{\mathcal{U}}^{\bot}$ is a GES, so $\mathcal{H}_{\mathcal{G}_{1}}\backslash \mathcal{H}_{\{|S\rangle\}}$ is also. For subspaces $\mathcal{H}_{\mathcal{U}}^{\bot}$ and $\mathcal{H}_{\mathcal{G}_{1}}\backslash \mathcal{H}_{\{|S\rangle\}}$, we gain the conclusion as below.

\emph{Theorem 3}. The states in GES $\mathcal{H}_{\mathcal{U}}^{\bot}$ is distillable across some bipartite cut, while the states belonging to GES $\mathcal{H}_{\mathcal{G}_{1}}\backslash \mathcal{H}_{\{|S\rangle\}}$ is distillable across every bipartition.

$Proof$. Suppose $\mathcal{P}(n)$ is the rank-$n$ projector ($1\leq n\leq 8$) acting on $\mathcal{H}_{\mathcal{U}}^{\bot}$. Based on the proof of Theorem 1, we know that any state in $\mathcal{H}_{\mathcal{U}}^{\bot}$ must be expressed as a linear combination of states from set $\mathcal{G}$ and constructing $n$ mutually orthogonal quantum states in $\mathcal{H}_{\mathcal{U}}^{\bot}$ requires at least $(n+1)$ states from $\mathcal{G}$. In $1|234$ bipartition, we calculate the reduced density matrix of every state in $\mathcal{G}$.
\begin{equation*}
\begin{aligned}
&\rho_{234}^{U1}=\textrm{Tr}_{1}(|\psi_{+}\rangle^{U1}\langle\psi_{+}|)=2|\varphi_1\rangle^{U1}\langle \varphi_1|+|\varphi_2\rangle^{U1}\langle \varphi_2|,\\
&\quad~~~~~~~~(|\varphi_1\rangle^{U1}=|000\rangle,|\varphi_2\rangle^{U1}=|0+1\rangle|00\rangle),\\
&\rho_{234}^{U2}=2|\varphi_1\rangle^{U2}\langle \varphi_1|+|\varphi_2\rangle^{U2}\langle \varphi_2|,\\
&\quad~~~~~~~~(|\varphi_1\rangle^{U2}=|222\rangle,|\varphi_2\rangle^{U2}=|1+2\rangle|22\rangle),\\
&\rho_{234}^{U3}=|\varphi_1\rangle^{U3}\langle \varphi_1|,(|\varphi_1\rangle^{U3}=|0\rangle|12+22+01+02\rangle),\\
&\rho_{234}^{U4}=|\varphi_1\rangle^{U4}\langle \varphi_1|,(|\varphi_1\rangle^{U4}=|2\rangle|00+10+20+21\rangle),\\
&\rho_{234}^{U5}=|\varphi_1\rangle^{U5}\langle \varphi_1|+|\varphi_2\rangle^{U5}\langle \varphi_2|,\\
&\quad~~~~~~~~(|\varphi_1\rangle^{U5}=|1+2\rangle|0+1\rangle|0\rangle,\\
&\quad\quad\quad~|\varphi_2\rangle^{U5}=|1+2\rangle|0+1\rangle|0+1+2\rangle),\\
&\rho_{234}^{U6}=|\varphi_1\rangle^{U6}\langle \varphi_1|,\\
&\quad~~~~~~~~(|\varphi_1\rangle^{U6}=|12+22+01+02\rangle|0+1\rangle),\\
\end{aligned}
\end{equation*}\\
\begin{equation*}
\begin{aligned}
&\rho_{234}^{U7}=|\varphi_1\rangle^{U7}\langle \varphi_1|+|\varphi_2\rangle^{U7}\langle \varphi_2|,\\
&\quad~~~~~~~~(|\varphi_1\rangle^{U7}=|0+1\rangle|1+2\rangle|2\rangle,\\
&\quad\quad\quad~|\varphi_2\rangle^{U7}=|0+1\rangle|1+2\rangle|0+1+2\rangle),\\
&\rho_{234}^{U8}=|\varphi_1\rangle^{U8}\langle \varphi_1|,\\
&\quad~~~~~~~~(|\varphi_1\rangle^{U8}=|00+10+20+21\rangle|1+2\rangle),\\
&\rho_{234}^{U9}=|\varphi_1\rangle^{U9}\langle \varphi_1|,(|\varphi_1\rangle^{U9}=|111\rangle).\\
\end{aligned}
\end{equation*}

We choose pure state $|\varphi_1\rangle^{Ui}$ in each $\rho_{234}^{Ui}$ with $i=1,2,\ldots,9$. They are linearly independent, because the rank of the matrix composed of these 9 column vectors is equal to 9. This means that each state in $\mathcal{G}$ contributes at least one independent rank in the bimarginal. So, $\mathcal{R}(\textrm{Tr}_{1}(\mathcal{P}(n)))\geq n+1>\mathcal{R}(\mathcal{P}(n))$. According to the Lemma 2, any genuinely entangled state in $\mathcal{H}_{\mathcal{U}}^{\bot}$ is 1-distillable in the $1|234$ cut. Similarly, they are also 1-distillable in the $2|341$, $3|412$, and $4|123$ cuts, but uncertain in $12|34$, $23|41$, and $13|24$ cuts.

$\mathcal{H}_{\mathcal{G}_{1}}\backslash \mathcal{H}_{\{|S\rangle\}}$ is a proper subspace of $\mathcal{H}_{\mathcal{U}}^{\bot}$. So, in $1|234$, $2|341$, $3|412$, and $4|123$ cuts, the states in $\mathcal{H}_{\mathcal{G}_{1}}\backslash \mathcal{H}_{\{|S\rangle\}}$ are also 1-distillable. In other cuts, we still use the above method and can find that each state in $\mathcal{G}_{1}$ contributes at least one independent rank in the corresponding bimarginal. For the projector $\mathcal{P}'(n)$ of rank-$n$ ($1\leq n\leq 6$) on subspace $\mathcal{H}_{\mathcal{G}_{1}}\backslash \mathcal{H}_{\{|S\rangle\}}$, we have $\mathcal{R}(\textrm{Tr}_{\ast}(\mathcal{P}'(n)))\geq n+1>\mathcal{R}(\mathcal{P}'(n))$. Here $\ast=12,23,13$. Therefore, all the states supported on $\mathcal{H}_{\mathcal{G}_{1}}\backslash \mathcal{H}_{\{|S\rangle\}}$ are distillable across every bipartition.  ~ \hfill $\square$

\section{Strongly nonlocal UBB in 4-qudit quantum system}\label{Q4}
We generalize the structure of the set $\mathcal{U}$ on 4-qutrit quantum system to 4-qudit quantum system. Fix an integer $d\geq3$. The computational basis of $(\mathbb{C}^{d})^{\otimes 4}$ quantum system is denoted as $\mathcal{B}=\{|i_{1}i_{2}i_{3}i_{4}\rangle\}=\{|i_{1}\rangle|i_{2}\rangle|i_{3}\rangle|i_{4}\rangle\}$ with $i_{1},i_{2},i_{3},i_{4}=0,1,\ldots,d-1$. For any integer $1\leq l\leq \lfloor\frac{d-1}{2}\rfloor$, we have the computational basis $\mathcal{B}^{l}=\bigcup_{i_{1},i_{2},i_{3},i_{4}=l-1}^{d-l}\{|i_{1}i_{2}i_{3}i_{4}\rangle\}\setminus \bigcup_{i_{1},i_{2},i_{3},i_{4}=l}^{d-l-1}\{|i_{1}i_{2}i_{3}i_{4}\rangle\}$ of $l$-th layer space of $(\mathbb{C}^{d})^{\otimes 4}$. Let $k=l-1$. We construct some orthogonal product sets and quantum states of the $l$-th layer space as follows.
\\
\begin{widetext}
The orthogonal product sets are expressed by
\begin{equation}\label{yel}
\begin{aligned}
&\mathcal{C}_{1}^{d,l}:=\{|\eta_{j_{1}}^{(d-2k)}\rangle_{1}|k\rangle_{2}|k\rangle_{3}|k\rangle_{4}~|~(j_{1})\in \textrm{Z}_{d-2k-1}\setminus \{(0)\}\},\\
&\mathcal{C}_{2}^{d,l}:=\{|k\rangle_{1}|\xi_{j_{2}}^{(d-2k)}\rangle_{2}|d-l\rangle_{3}|d-l\rangle_{4}~|~(j_{2})\in \textrm{Z}_{d-2k-1}\setminus \{(0)\}\},\\
&\mathcal{C}_{3}^{d,l}:=\{|k\rangle_{1}|k\rangle_{2}|\xi_{j_{3}}^{(d-2k)}\rangle_{3}|d-l\rangle_{4}~|~(j_{3})\in \textrm{Z}_{d-2k-1}\setminus \{(0)\}\},\\
&\mathcal{C}_{4}^{d,l}:=\{|k\rangle_{1}|k\rangle_{2}|k\rangle_{3}|\xi_{j_{4}}^{(d-2k)}\rangle_{4}~|~(j_{4})\in \textrm{Z}_{d-2k-1}\setminus \{(0)\}\},\\
&\mathcal{C}_{5}^{d,l}:=\{|\eta_{j_{1}}^{(d-2k)}\rangle_{1}|\xi_{j_{2}}^{(d-2k)}\rangle_{2}|\eta_{j_{3}}^{(d-2k)}\rangle_{3}|k\rangle_{4}~|~(j_{1},j_{2},j_{3})\in \textrm{Z}_{d-2k-1}\times \textrm{Z}_{d-2k-1}\times \textrm{Z}_{d-2k-1}\setminus \{(0,0,0)\}\},\\
&\mathcal{C}_{6}^{d,l}:=\{|\eta_{j_{1}}^{(d-2k)}\rangle_{1}|\xi_{j_{2}}^{(d-2k)}\rangle_{2}|d-l\rangle_{3}|\eta_{j_{4}}^{(d-2k)}\rangle_{4}~|~(j_{1},j_{2},j_{4})\in \textrm{Z}_{d-2k-1}\times \textrm{Z}_{d-2k-1}\times \textrm{Z}_{d-2k-1}\setminus \{(0,0,0)\}\},\\
\end{aligned}
\end{equation}
\begin{equation*}
\begin{aligned}
&\mathcal{C}_{7}^{d,l}:=\{|\eta_{j_{1}}^{(d-2k)}\rangle_{1}|k\rangle_{2}|\xi_{j_{3}}^{(d-2k)}\rangle_{3}|\eta_{j_{4}}^{(d-2k)}\rangle_{4}~|~(j_{1},j_{3},j_{4})\in \textrm{Z}_{d-2k-1}\times \textrm{Z}_{d-2k-1}\times \textrm{Z}_{d-2k-1}\setminus \{(0,0,0)\}\},\\
&\mathcal{C}_{8}^{d,l}:=\{|k\rangle_{1}|\xi_{j_{2}}^{(d-2k)}\rangle_{2}|\eta_{j_{3}}^{(d-2k)}\rangle_{3}|\xi_{j_{4}}^{(d-2k)}\rangle_{4}~|~(j_{2},j_{3},j_{4})\in \textrm{Z}_{d-2k-1}\times \textrm{Z}_{d-2k-1}\times \textrm{Z}_{d-2k-1}\setminus \{(0,0,0)\}\},\\
&\mathcal{D}_{1}^{d,l}:=\{|\xi_{j_{1}}^{(d-2k)}\rangle_{1}|d-l\rangle_{2}|d-l\rangle_{3}|d-l\rangle_{4}~|~(j_{1})\in \textrm{Z}_{d-2k-1}\setminus \{(0)\}\},\\
&\mathcal{D}_{2}^{d,l}:=\{|d-l\rangle_{1}|\eta_{j_{2}}^{(d-2k)}\rangle_{2}|k\rangle_{3}|k\rangle_{4}~|~(j_{2})\in \textrm{Z}_{d-2k-1}\setminus \{(0)\}\},\\
&\mathcal{D}_{3}^{d,l}:=\{|d-l\rangle_{1}|d-l\rangle_{2}|\eta_{j_{3}}^{(d-2k)}\rangle_{3}|k\rangle_{4}~|~(j_{3})\in \textrm{Z}_{d-2k-1}\setminus \{(0)\}\},\\
&\mathcal{D}_{4}^{d,l}:=\{|d-l\rangle_{1}|d-l\rangle_{2}|d-l\rangle_{3}|\eta_{j_{4}}^{(d-2k)}\rangle_{4}~|~(j_{4})\in \textrm{Z}_{d-2k-1}\setminus \{(0)\}\},\\
&\mathcal{D}_{5}^{d,l}:=\{|\xi_{j_{1}}^{(d-2k)}\rangle_{1}|\eta_{j_{2}}^{(d-2k)}\rangle_{2}|\xi_{j_{3}}^{(d-2k)}\rangle_{3}|d-l\rangle_{4}~|~(j_{1},j_{2},j_{3})\in \textrm{Z}_{d-2k-1}\times \textrm{Z}_{d-2k-1}\times \textrm{Z}_{d-2k-1}\setminus \{(0,0,0)\}\},\\
&\mathcal{D}_{6}^{d,l}:=\{|\xi_{j_{1}}^{(d-2k)}\rangle_{1}|\eta_{j_{2}}^{(d-2k)}\rangle_{2}|k\rangle_{3}|\xi_{j_{4}}^{(d-2k)}\rangle_{4}~|~(j_{1},j_{2},j_{4})\in \textrm{Z}_{d-2k-1}\times \textrm{Z}_{d-2k-1}\times \textrm{Z}_{d-2k-1}\setminus \{(0,0,0)\}\},\\
&\mathcal{D}_{7}^{d,l}:=\{|\xi_{j_{1}}^{(d-2k)}\rangle_{1}|d-l\rangle_{2}|\eta_{j_{3}}^{(d-2k)}\rangle_{3}|\xi_{j_{4}}^{(d-2k)}\rangle_{4}~|~(j_{1},j_{3},j_{4})\in \textrm{Z}_{d-2k-1}\times \textrm{Z}_{d-2k-1}\times \textrm{Z}_{d-2k-1}\setminus \{(0,0,0)\}\},\\
\end{aligned}
\end{equation*}

\begin{equation*}
\begin{aligned}
&\mathcal{D}_{8}^{d,l}:=\{|d-l\rangle_{1}|\eta_{j_{2}}^{(d-2k)}\rangle_{2}|\xi_{j_{3}}^{(d-2k)}\rangle_{3}|\eta_{j_{4}}^{(d-2k)}\rangle_{4}~|~(j_{2},j_{3},j_{4})\in \textrm{Z}_{d-2k-1}\times \textrm{Z}_{d-2k-1}\times \textrm{Z}_{d-2k-1}\setminus \{(0,0,0)\}\}.\\
\end{aligned}
\end{equation*}

Define some quantum states as
\begin{equation}\label{yels}
\begin{aligned}
&|\psi_{\pm}\rangle_{1}^{d,l}=|\eta_{0}^{(d-2k)}\rangle_{1}|k\rangle_{2}|k\rangle_{3}|k\rangle_{4}\pm |d-l\rangle_{1}|\eta_{0}^{(d-2k)}\rangle_{2}|k\rangle_{3}|k\rangle_{4},\\
&|\psi_{\pm}\rangle_{2}^{d,l}=|k\rangle_{1}|\xi_{0}^{(d-2k)}\rangle_{2}|d-l\rangle_{3}|d-l\rangle_{4}\pm |\xi_{0}^{(d-2k)}\rangle_{1}|d-l\rangle_{2}|d-l\rangle_{3}|d-l\rangle_{4},\\
&|\psi_{\pm}\rangle_{3}^{d,l}=\{|k\rangle_{1}|k\rangle_{2}|\xi_{0}^{(d-2k)}\rangle_{3}|d-l\rangle_{4}\pm |k\rangle_{1}|k\rangle_{2}|k\rangle_{3}|\xi_{0}^{(d-2k)}\rangle_{4},\\
&|\psi_{\pm}\rangle_{4}^{d,l}=|d-l\rangle_{1}|d-l\rangle_{2}|\eta_{0}^{(d-2k)}\rangle_{3}|k\rangle_{4}\pm |d-l\rangle_{1}|d-l\rangle_{2}|d-l\rangle_{3}|\eta_{0}^{(d-2k)}\rangle_{4},\\
&|\psi_{\pm}\rangle_{5}^{d,l}=|\eta_{0}^{(d-2k)}\rangle_{1}|\xi_{0}^{(d-2k)}\rangle_{2}|\eta_{0}^{(d-2k)}\rangle_{3}|k\rangle_{4}\pm |k\rangle_{1}|\xi_{0}^{(d-2k)}\rangle_{2}|\eta_{0}^{(d-2k)}\rangle_{3}|\xi_{0}^{(d-2k)}\rangle_{4},\\
&|\psi_{\pm}\rangle_{6}^{d,l}=|\eta_{0}^{(d-2k)}\rangle_{1}|\xi_{0}^{(d-2k)}\rangle_{2}|d-l\rangle_{3}|\eta_{0}^{(d-2k)}\rangle_{4}\pm |\eta_{0}^{(d-2k)}\rangle_{1}|k\rangle_{2}|\xi_{0}^{(d-2k)}\rangle_{3}|\eta_{0}^{(d-2k)}\rangle_{4},\\
&|\psi_{\pm}\rangle_{7}^{d,l}=|\xi_{0}^{(d-2k)}\rangle_{1}|\eta_{0}^{(d-2k)}\rangle_{2}|\xi_{0}^{(d-2k)}\rangle_{3}|d-l\rangle_{4}\pm |d-l\rangle_{1}|\eta_{0}^{(d-2k)}\rangle_{2}|\xi_{0}^{(d-2k)}\rangle_{3}|\eta_{0}^{(d-2k)}\rangle_{4},\\
&|\psi_{\pm}\rangle_{8}^{d,l}=|\xi_{0}^{(d-2k)}\rangle_{1}|\eta_{0}^{(d-2k)}\rangle_{2}|k\rangle_{3}|\xi_{0}^{(d-2k)}\rangle_{4}\pm |\xi_{0}^{(d-2k)}\rangle_{1}|d-l\rangle_{2}|\eta_{0}^{(d-2k)}\rangle_{3}|\xi_{0}^{(d-2k)}\rangle_{4}.\\
\end{aligned}
\end{equation}
Here $|\eta_{j}^{(d-2k)}\rangle=\sum_{t=k}^{d-k-2}\omega_{d-2k-1}^{j(t-k)}|t\rangle$ and $|\xi_{j}^{(d-2k)}\rangle=\sum_{t=k}^{d-k-2}\omega_{d-2k-1}^{j(t-k)}|t+1\rangle$ for $j\in \textrm{Z}_{d-2k-1}$, where $\omega_{d-2k-1}:=\textrm{e}^{\frac{2\pi\sqrt{-1}}{d-2k-1}}$.

Specially, when $d$ is even, we define
\begin{equation}\label{yele}
\begin{aligned}
\mathcal{C}^{d,d/2}:=\{|\phi_{j_{1}}\rangle_{1}|\phi_{j_{2}}\rangle_{2}|\phi_{j_{3}}\rangle_{3}|\phi_{j_{4}}\rangle_{4}~|~(j_{1},j_{2},j_{3},j_{4})\in \textrm{Z}_{2}\times \textrm{Z}_{2}\times \textrm{Z}_{2}\times \textrm{Z}_{2}\setminus \{(0,0,0,0)\}\},
\end{aligned}
\end{equation}
where $|\phi_{j}\rangle=|\frac{d-2}{2}\rangle+(-1)^j|\frac{d}{2}\rangle$ with $j\in \textrm{Z}_{2}$.
\end{widetext}

In $(\mathbb{C}^{d})^{\otimes 4}$ quantum system, the stopper state is $|S\rangle^{d}=\left(\sum_{i_{1}=0}^{d-1}|i_{1}\rangle\right)\bigotimes\left(\sum_{i_{2}=0}^{d-1}|i_{2}\rangle\right)\bigotimes \left(\sum_{i_{3}=0}^{d-1}|i_{3}\rangle\right)\bigotimes$ $\left(\sum_{i_{4}=0}^{d-1}|i_{4}\rangle\right)$. Let
\begin{equation*}
\begin{aligned}
&\mathcal{U}^{d,l}_{1,-}=\mathcal{C}_{1}^{d,l}\cup \mathcal{D}_{2}^{d,l}\cup |\psi_{-}\rangle_{1}^{d,l},\\
&\mathcal{U}^{d,l}_{2,-}=\mathcal{C}_{2}^{d,l}\cup \mathcal{D}_{1}^{d,l}\cup |\psi_{-}\rangle_{2}^{d,l},\\
&\mathcal{U}^{d,l}_{3,-}=\mathcal{C}_{3}^{d,l}\cup \mathcal{C}_{4}^{d,l}\cup |\psi_{-}\rangle_{3}^{d,l},\\
&\mathcal{U}^{d,l}_{4,-}=\mathcal{D}_{3}^{d,l}\cup \mathcal{D}_{4}^{d,l}\cup |\psi_{-}\rangle_{4}^{d,l},\\
&\mathcal{U}^{d,l}_{5,-}=\mathcal{C}_{5}^{d,l}\cup \mathcal{C}_{8}^{d,l}\cup |\psi_{-}\rangle_{5}^{d,l},\\
&\mathcal{U}^{d,l}_{6,-}=\mathcal{C}_{6}^{d,l}\cup \mathcal{C}_{7}^{d,l}\cup |\psi_{-}\rangle_{6}^{d,l},\\
&\mathcal{U}^{d,l}_{7,-}=\mathcal{D}_{5}^{d,l}\cup \mathcal{D}_{8}^{d,l}\cup |\psi_{-}\rangle_{7}^{d,l},\\
&\mathcal{U}^{d,l}_{8,-}=\mathcal{D}_{6}^{d,l}\cup \mathcal{D}_{7}^{d,l}\cup |\psi_{-}\rangle_{8}^{d,l}.\\
\end{aligned}
\end{equation*}

\emph{Theorem 4}. In $(\mathbb{C}^{d})^{\otimes 4}$, $d\geq 3$, the set
\begin{equation}\label{ud}
\begin{aligned}\mathcal{U}^{d}=\begin{cases}
\left(\bigcup_{i,l}\mathcal{U}^{d,l}_{i,-}\right)\bigcup\{|S\rangle^{d}\}, &d~\textrm{is~odd}\\
~\\
\left(\bigcup_{i,l}\mathcal{U}^{d,l}_{i,-}\right)\bigcup\mathcal{C}^{d,d/2}\bigcup\{|S\rangle^{d}\}, &d~\textrm{is~even}
\end{cases}
\end{aligned}
\end{equation}
is a UBB with strong quantum nonlocality.

$Proof$. It is obvious that all the quantum states belonging to the set $\mathcal{U}^{d}$ are biseparable and pairwise orthogonal. We need to demonstrate its unextendibility. The method similar to the proof of Theorem 1. When $d$ is odd, the states $|\frac{d-1}{2}\rangle^{\otimes 4}$ and $|\psi_{+}\rangle_{i}^{d,l}$ ($i=1,\ldots,8$ and $l=1,\ldots,\lfloor\frac{d-1}{2}\rfloor$) are not orthogonal to $|S\rangle^{d}$ but are orthogonal to all states in $\mathcal{U}^{d}\setminus|S\rangle^{d}$, the any state $|\phi\rangle\in\mathcal{H}_{\mathcal{U}^{d}}^{\bot}$ can be written as a linear combination of above states such that $\langle \phi|S\rangle^{d}=0$.

\begin{widetext}~
\begin{figure}[h]
\centering
\includegraphics[width=0.95\textwidth]{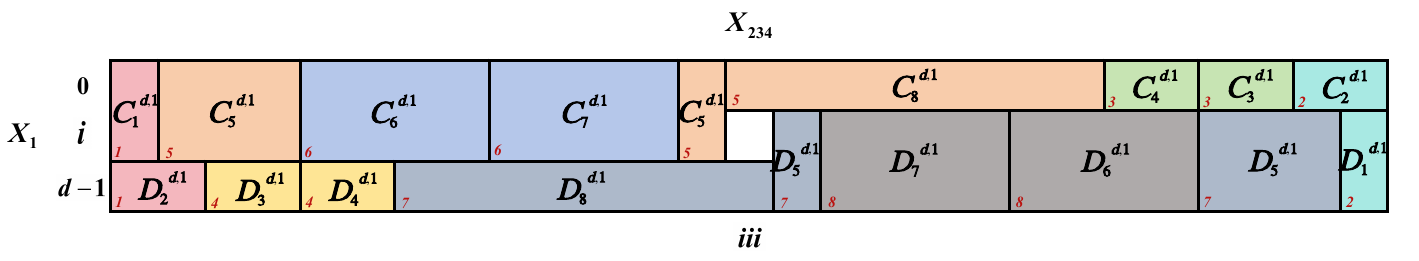}
\caption{This is the structure of set $\mathcal{U}^{d}$ under bipartition $X_{1}|X_{234}$. In figure, $i$ expresses the dimension of $1,\ldots,d-2$. The regions denoted as $\mathcal{C}_{j}^{d,1}$ and $\mathcal{D}_{j}^{d,1}$ with $j=1,\ldots,8$ correspond to the subsets of Eq. (\ref{yel}), respectively. Each colored area is marked with a red number in the left lower corner. The union of the regions with the same color and red number $j$ correspond the subsets $\mathcal{U}_{j}^{-}$ of set $\mathcal{U}^{d}$. \label{Tu3}}
\end{figure}

Specifically,
\begin{equation*}
|\phi\rangle=p|\frac{d-1}{2}\rangle^{\otimes 4}+\sum_{l=1}^{\lfloor\frac{d-1}{2}\rfloor}(a_{l}|\psi_{+}\rangle_{1}^{d,l}+b_{l}|\psi_{+}\rangle_{2}^{d,l}+c_{l}|\psi_{+}\rangle_{3}^{d,l}+d_{l}|\psi_{+}\rangle_{4}^{d,l}+e_{l}|\psi_{+}\rangle_{5}^{d,l}+ f_{l}|\psi_{+}\rangle_{6}^{d,l}+g_{l}|\psi_{+}\rangle_{7}^{d,l}+h_{l}|\psi_{+}\rangle_{8}^{d,l}),
\end{equation*}
where the coefficients have at least two non-zero values. Assume $|\phi\rangle$ is separable in $1|234$ bipartition. Then, it corresponds to a matrix
\begin{equation*}
\begin{aligned}
M_1^{d}=\left(\begin{array}{ccccccccccccccccccccccccccccc}
a_{1} & c_{1} & c_{1} & f_{1} & f_{1} & c_{1} & f_{1} & f_{1} & c_{1} & e_{1} & e_{1} & e_{1} & e_{1} & e_{1} & \cdots & e_{1} & e_{1} & f_{1} & f_{1} & b_{1} & e_{1} & e_{1} & e_{1} & e_{1} & e_{1} & e_{1} & f_{1} & f_{1} & b_{1} \\
a_{1} & h_{1} & h_{1} & f_{1} & f_{1} & g_{1} & f_{1} & f_{1} & g_{1} & e_{1} & h_{1} & h_{1} & e_{1} &\ulcorner&        &\urcorner& g_{1} & f_{1} & f_{1} & g_{1} & e_{1} & h_{1} & h_{1} & e_{1} & h_{1} & h_{1} & f_{1} & f_{1} & b_{1} \\
\vdots & \vdots & \vdots & \vdots & \vdots & \vdots & \vdots & \vdots & \vdots & \vdots & \vdots & \vdots & \vdots &   &M_{2}^{d}&   & \vdots & \vdots & \vdots & \vdots & \vdots & \vdots & \vdots & \vdots & \vdots & \vdots & \vdots & \vdots & \vdots \\
a_{1} & h_{1} & h_{1} & f_{1} & f_{1} & g_{1} & f_{1} & f_{1} & g_{1} & e_{1} & h_{1} & h_{1} & e_{1} &\llcorner&        &\lrcorner& g_{1} & f_{1} & f_{1} & g_{1} & e_{1} & h_{1} & h_{1} & e_{1} & h_{1} & h_{1} & f_{1} & f_{1} & b_{1} \\
a_{1} & h_{1} & h_{1} & g_{1} & g_{1} & g_{1} & g_{1} & g_{1} & g_{1} & a_{1} & h_{1} & h_{1} & g_{1} & g_{1} & \cdots & g_{1} & g_{1} & g_{1} & g_{1} & g_{1} & d_{1} & h_{1} & h_{1} & d_{1} & h_{1} & h_{1} & d_{1} & d_{1} & b_{1} \\
\end{array}
\right).
\end{aligned}
\end{equation*}
It is the $d\times d^2$ coefficient matrix of the $1$-th layer and has the same structure as $M_{1}$ in the proof of Theorem 1. The center part $M_2^{d}$ is the coefficient matrix of the $2$-th layer similar to $M_1^{d}$. We iterate in this manner until the coefficient matrix
\begin{equation*}
\begin{aligned}
M_o^{d}=\left(\begin{array}{ccccccccccccccccccccccccccccc}
a_{o} & c_{o} & c_{o} & f_{o} & f_{o} & c_{o} & f_{o} & f_{o} & c_{o} & e_{o} & e_{o} & e_{o} & e_{o} & e_{o} & e_{o} & f_{o} & f_{o} & b_{o} & e_{o} & e_{o} & e_{o} & e_{o} & e_{o} & e_{o} & f_{o} & f_{o} & b_{o} \\
a_{o} & h_{o} & h_{o} & f_{o} & f_{o} & g_{o} & f_{o} & f_{o} & g_{o} & e_{o} & h_{o} & h_{o} & e_{o} &    p  & g_{o} & f_{o} & f_{o} & g_{o} & e_{o} & h_{o} & h_{o} & e_{o} & h_{o} & h_{o} & f_{o} & f_{o} & b_{o} \\
a_{o} & h_{o} & h_{o} & g_{o} & g_{o} & g_{o} & g_{o} & g_{o} & g_{o} & a_{o} & h_{o} & h_{o} & g_{o} & g_{o} & g_{o} & g_{o} & g_{o} & g_{o} & d_{o} & h_{o} & h_{o} & d_{o} & h_{o} & h_{o} & d_{o} & d_{o} & b_{o} \\
\end{array}
\right)
\end{aligned}
\end{equation*}
of the $\lfloor\frac{d-1}{2}\rfloor$-th layer. Here $o=\lfloor\frac{d-1}{2}\rfloor$.
\end{widetext}
Meanwhile $\textrm{rank}(M_1^{d})=1$. According to the condition of at least two non-zero coefficients and the result in the proof of Theorem 1, we can be deduce that all coefficients are equal and non-zero. From this, we have $\langle \phi|S\rangle^{d}\neq 0$. This is contradictory to the orthogonality between $|\phi\rangle$ and $|S\rangle^{d}$. Thus, $|\phi\rangle$ is inseparable in $1|234$ bipartition. By using the same way, it can be inferred that $|\phi\rangle$ is also inseparable in other bipartition. In other words, any state in subspace $\mathcal{H}_{\mathcal{U}^{d}}^{\bot}$ is genuinely entangled.

When $d$ is even, we can obtain the same conclusion by simply substituting $|\frac{d-1}{2}\rangle^{\otimes 4}$ with $\left(|\frac{d-2}{2}\rangle+|\frac{d}{2}\rangle\right)^{\otimes 4}$ in the process of proof. So, $\mathcal{U}^{d}$ is indeed a UBB.

On the other hand, the union $\left(\bigcup_{i,1}\mathcal{U}^{d,1}_{i,-}\right)\bigcup\{|S\rangle^{d}\}$ of the quantum states in $1$-th layer and stopper state have same structure with the strongly nonlocal UBB $\mathcal{U}$ as shown in the Fig. \ref{Tu3}. It is not difficult to deduce that the subset $\left(\bigcup_{i,1}\mathcal{U}^{d,1}_{i,-}\right)\bigcup\{|S\rangle^{d}\}$ have the strong quantum nonlocality by using the manner in proof of Theorem 2. Therefore, the complete set $\mathcal{U}^{d}$ also possesses strong quantum nonlocality.
~ \hfill $\square$

According to the Theorem 4, the complementary subspace of $\mathcal{H}_{\mathcal{U}^{d}}$ contains only genuinely entangled states and $\dim(\mathcal{H}_{\mathcal{U}^{d}}^{\perp})=8\lfloor\frac{d-1}{2}\rfloor$. Define quantum states \begin{equation}\label{gef1}
\begin{aligned}
&~~~~~~|F\rangle^{d,l}\\
&=\frac{1}{4[(d-2k-1)^2+1]\sqrt{2(d-2k-1)}}\left(\sum_{i=1}^{4}|\widetilde{\psi_{+}}\rangle_{i}^{d,l}\right)\\
&~~~~~~~~~~~~~~+\frac{\sqrt{(d-2k-1)}}{4\sqrt{2}[(d-2k-1)^2+1]}\left(\sum_{i=5}^{8}|\widetilde{\psi_{+}}\rangle_{i}^{d,l}\right),\\
\end{aligned}
\end{equation}
for $l=1,2,\ldots,\lfloor\frac{d-1}{2}\rfloor$. When $d$ is odd, for $l=\frac{d+1}{2}$, we have the center state $|F\rangle^{d,(d+1)/2}=|\frac{d-1}{2}\rangle^{\otimes 4}$. When $d$ is even, for $l=\frac{d}{2}$, we have the center state $|F\rangle^{d,d/2}=\frac{1}{16}\left(|\frac{d-2}{2}\rangle+|\frac{d}{2}\rangle\right)^{\otimes 4}$. The inner products of these states and $|S\rangle^{d}$ are all equal to 1. Then, the orthonormal basis of $\mathcal{H}_{\mathcal{U}^{d}}^{\perp}$ can be expressed as the union of the following states
\begin{equation}\label{ges1}
\begin{aligned}
&|G_1\rangle^{d,l}=\frac{1}{\sqrt{2}}\left(|\widetilde{\psi_{+}}\rangle_{1}^{d,l}-|\widetilde{\psi_{+}}\rangle_{2}^{d,l}\right),\\
&|G_2\rangle^{d,l}=\frac{1}{\sqrt{2}}\left(|\widetilde{\psi_{+}}\rangle_{3}^{d,l}-|\widetilde{\psi_{+}}\rangle_{4}^{d,l}\right),\\
&|G_3\rangle^{d,l}=\frac{1}{2}\left(|\widetilde{\psi_{+}}\rangle_{1}^{d,l}+|\widetilde{\psi_{+}}\rangle_{2}^{d,l}-|\widetilde{\psi_{+}}\rangle_{3}^{d,l}-|\widetilde{\psi_{+}}\rangle_{4}^{d,l}\right),\\
&|G_4\rangle^{d,l}=\frac{1}{\sqrt{2}}\left(|\widetilde{\psi_{+}}\rangle_{5}^{d,l}-|\widetilde{\psi_{+}}\rangle_{6}^{d,l}\right),\\
&|G_5\rangle^{d,l}=\frac{1}{\sqrt{2}}\left(|\widetilde{\psi_{+}}\rangle_{7}^{d,l}-|\widetilde{\psi_{+}}\rangle_{8}^{d,l}\right),\\
&|G_6\rangle^{d,l}=\frac{1}{2}\left(|\widetilde{\psi_{+}}\rangle_{5}^{d,l}+|\widetilde{\psi_{+}}\rangle_{6}^{d,l}-|\widetilde{\psi_{+}}\rangle_{7}^{d,l}-|\widetilde{\psi_{+}}\rangle_{8}^{d,l}\right),\\
&|G_7\rangle^{d,l}=\frac{d-2k-1}{\sqrt{4(d-2k-1)^2+4}}\left(\sum_{i=1}^4|\widetilde{\psi_{+}}\rangle_{i}^{d,l}\right)\\
&~~~~~~~~~~~~~~~~-\frac{1}{\sqrt{4(d-2k-1)^2+4}}\left(\sum_{i=5}^8|\widetilde{\psi_{+}}\rangle_{i}^{d,l}\right),\\
&|G_8\rangle^{d,l}=\sum_{t=0}^{\lfloor\frac{d-1}{2}\rfloor}\omega_{\lfloor\frac{d+1}{2}\rfloor}^{lt}|F\rangle^{d,t+1}\Bigg/\sqrt{\sum_{t=0}^{\lfloor\frac{d-1}{2}\rfloor}\langle F|F\rangle^{d,t+1}},\\
\end{aligned}
\end{equation}
for $l=1,2,\ldots,\lfloor\frac{d-1}{2}\rfloor$. Here $\langle F|F\rangle^{d,t+1}=\{8[(d-2t-1)^2+1](d-2t-1)\}^{-1}$ with $t=0,\ldots,\lfloor\frac{d-3}{2}\rfloor$. When $d$ is odd, there is $\langle F|F\rangle^{d,\lfloor\frac{d+1}{2}\rfloor}=1$. When $d$ is even, there is $\langle F|F\rangle^{d,\lfloor\frac{d+1}{2}\rfloor}=\frac{1}{16}$.

Due to the same structure between every layer subsets of $\mathcal{U}^{d}$ and $\mathcal{U}$, the reduced density operators of states $|S\rangle^{d}$ and $|\psi_{+}\rangle_{i}^{d,l}$ $(i=1,\ldots,8,l=1,\ldots,\lfloor\frac{d-1}{2}\rfloor)$ for systems $X_{234}$, $X_{341}$, $X_{412}$, and $X_{123}$ contribute at least one independent rank, respectively. Moreover, the state $|\psi_{+}\rangle_{i}^{d,l}$ $(i=1,\ldots,7,l=1,\ldots,\lfloor\frac{d-1}{2}\rfloor)$ contributes at least one independent rank in the bimarginal of bipartitions $12|34$, $23|41$, and $13|24$, respectively. Let $\mathcal{G}_{d1}=\{|\psi_{+}\rangle_{i}^{d,l}\}_{i,l=1,1}^{7,\lfloor\frac{d-1}{2}\rfloor}$. So, for $\mathcal{H}_{\mathcal{U}^{d}}^{\perp}$ and $\mathcal{H}_{\mathcal{G}_{d1}}\backslash \mathcal{H}_{\{|S\rangle^{d}\}}$, we can directly deduce the following conclusion.

\emph{Theorem 5}. In $(\mathbb{C}^{d})^{\otimes 4}$ $(d\geq 3)$ quantum system, the GES $\mathcal{H}_{\mathcal{U}^{d}}^{\bot}$ is distillable across some bipartite cut, while the GES $\mathcal{H}_{\mathcal{G}_{d1}}\backslash \mathcal{H}_{\{|S\rangle^{d}\}}$ is distillable across every bipartition.

The dimension of the subspace $\mathcal{H}_{\mathcal{G}_{d1}}\backslash \mathcal{H}_{\{|S\rangle^{d}\}}$ is $7\lfloor\frac{d-1}{2}\rfloor-1$, which is $\lfloor\frac{d-1}{2}\rfloor+1$ less than the dimension of $\mathcal{H}_{\mathcal{U}^{d}}^{\bot}$. The GES $\mathcal{H}_{\mathcal{G}_{d1}}\backslash \mathcal{H}_{\{|S\rangle^{d}\}}$ is a proper subspace of $\mathcal{H}_{\mathcal{U}^{d}}^{\bot}$. Although this subspace is smaller than $\mathcal{H}_{\mathcal{U}^{d}}^{\bot}$, the genuinely entangled states in this space are one-copy distillable under every bipartition.

\section{Conclusion}\label{Q5}
We have investigated the strongly nonlocal UBBs and the distillable GESs in four-partite quantum systems. In $\mathbb{C}^{3}\otimes \mathbb{C}^{3}\otimes \mathbb{C}^{3}\otimes \mathbb{C}^{3}$, by adjusting the structure of known OPS $\mathcal{E}$ with strong quantum nonlocality, we obtain a UBB $\mathcal{U}$ and prove that the UBB still retains the strong quantum nonlocality. The complementary subspace $\mathcal{H}_{\mathcal{U}}^{\bot}$ of subspace $\mathcal{H}_{\mathcal{U}}$ spanned by set $\mathcal{U}$ is a GES which contains only genuinely entangled states. We not only provide the specific form of the orthonormal basis for GES $\mathcal{H}_{\mathcal{U}}^{\bot}$, but also give a proper subspace $\mathcal{H}_{\mathcal{G}_{1}}\backslash \mathcal{H}_{\{|S\rangle\}}$ of $\mathcal{H}_{\mathcal{U}}^{\bot}$. The states in such proper subspace $\mathcal{H}_{\mathcal{G}_{1}}\backslash \mathcal{H}_{\{|S\rangle\}}$ are distillable across every bipartition. Based on the characteristics of planar structure of the newly constructed UBB $\mathcal{U}$, we generalize all results of each subsystem with dimension 3 to high-dimensional systems $\mathbb{C}^{d}\otimes \mathbb{C}^{d}\otimes \mathbb{C}^{d}\otimes \mathbb{C}^{d}$ $(d\geq 3)$. These will supply significant theoretical tools for the application research of quantum information.

\begin{acknowledgments}
This work was supported by the National Natural Science Foundation of China under grant nos. 12526564 and 62271189, the Natural Science Foundation of Hebei Province under grant no. A2025403008, and the Doctoral Science Start Foundation of Hebei GEO University of China under grant no. BQ2024075.
\end{acknowledgments}
~\\

\begin{appendix}
\begin{widetext}
\section{The matrixs corresponding to $|\phi\rangle$ in different bipartitions}\label{A}
In $2|341$ bipartition, the corresponding matrix is
\begin{equation*}
\begin{aligned}
M_2=\left(\begin{array}{ccccccccccccccccccccccccccc}
a & a & a & c & h & h & c & h & h & f & f & g & f & f & g & c & g & g & f & f & g & f & f & g & c & g & g\\
e & e & a & e & h & h & e & h & h & e & e & g & e & k & g & e & g & g & f & f & g & f & f & g & b & g & g\\
e & e & d & e & h & h & e & h & h & e & e & d & e & h & h & e & h & h & f & f & d & f & f & d & b & b & b\\
\end{array}
\right).
\end{aligned}
\end{equation*}

In $3|412$ bipartition, the corresponding matrix is
\begin{equation*}
\begin{aligned}
M_3=\left(\begin{array}{ccccccccccccccccccccccccccc}
a & e & e & a & e & e & a & a & d & c & e & e & h & h & h & h & h & h & c & e & e & h & h & h & h & h & h\\
f & e & e & f & e & e & g & g & d & f & e & e & f & k & h & g & g & h & c & e & e & g & g & h & g & g & h\\
f & f & f & f & f & f & g & g & d & f & f & f & f & f & f & g & g & d & c & b & b & g & g & b & g & g & b\\
\end{array}
\right).
\end{aligned}
\end{equation*}

In $4|123$ bipartition, the corresponding matrix is
\begin{equation*}
\begin{aligned}
M_4=\left(\begin{array}{ccccccccccccccccccccccccccc}
a & f & f & e & e & f & e & e & f & a & f & f & e & e & f & e & e & f & a & g & g & a & g & g & d & d & d\\
c & f & f & e & e & f & e & e & f & h & f & f & h & k & f & h & h & f & h & g & g & h & g & g & h & h & d\\
c & c & c & e & e & b & e & e & b & h & g & g & h & g & g & h & h & b & h & g & g & h & g & g & h & h & b\\
\end{array}
\right).
\end{aligned}
\end{equation*}

In $12|34$, $13|24$, and $14|23$ bipartitions, the corresponding matrix are
\begin{equation*}
\begin{aligned}
M_5=\left(\begin{array}{ccccccccc}
a & c & c & f & f & c & f & f & c\\
e & e & e & e & e & e & f & f & b\\
e & e & e & e & e & e & f & f & b\\
a & h & h & f & f & g & f & f & g\\
e & h & h & e & k & g & f & f & g\\
e & h & h & e & h & h & f & f & b\\
a & h & h & g & g & g & g & g & g\\
a & h & h & g & g & g & g & g & g\\
d & h & h & d & h & h & d & d & b\\
\end{array}
\right),
M_6=\left(\begin{array}{ccccccccc}
a & c & c & e & e & e & e & e & e\\
f & f & c & e & e & e & e & e & e\\
f & f & c & f & f & b & f & f & b\\
a & h & h & e & h & h & e & h & h\\
f & f & g & e & k & g & e & h & h\\
f & f & g & f & f & g & f & f & b\\
a & h & h & a & h & h & d & h & h\\
g & g & g & g & g & g & d & h & h\\
g & g & g & g & g & g & d & d & b\\
\end{array}
\right),
M_7=\left(\begin{array}{ccccccccc}
a & f & f & e & e & f & e & e & f\\
c & f & f & e & e & f & e & e & f\\
c & c & c & e & e & b & e & e & b\\
a & f & f & e & e & f & e & e & f\\
h & f & f & h & k & f & h & h & f\\
h & g & g & h & g & g & h & h & b\\
a & g & g & a & g & g & d & d & d\\
h & g & g & h & g & g & h & h & d\\
h & g & g & h & g & g & h & h & b\\
\end{array}
\right),
\end{aligned}
\end{equation*}
respectively.
\end{widetext}

\end{appendix}

\end{document}